\documentclass[manuscript,acmsmall,reviews,screen]{acmart}
\preto{\nolinenumbers} 

\AtBeginDocument{%
  }

\setcopyright{acmlicensed}
\copyrightyear{2025}
\acmYear{2025}

\acmJournal{JACM}
\acmVolume{37}
\acmNumber{4}
\acmArticle{111}
\acmMonth{8}

\usepackage{amsmath}
\usepackage{graphicx}
\usepackage{epstopdf}
\usepackage{textcomp}
\usepackage{xcolor}
\usepackage{listings}
\usepackage{wrapfig}

\usepackage{times}
\usepackage{subfigure} 
\usepackage{float}
\usepackage{booktabs}
\usepackage[linesnumbered,ruled,vlined]{algorithm2e}
\usepackage{setspace}
\usepackage{svg} 
\usepackage{etoolbox}
\usepackage{url}
\usepackage{newtxtext,newtxmath}
\usepackage{tikz}
\usepackage{tabularx}
\usepackage{booktabs}
\usepackage{multirow}
\usepackage{makecell}
\usepackage{array}
\usepackage{caption}
\usepackage{cancel}
\usepackage{colortbl}
\usepackage[normalem]{ulem}
\usepackage{verbatim}
\newcommand\name{MERE}
\usepackage{url}
\usepackage{subcaption}
\usepackage{pifont}
\usepackage{tabularx}     
\usepackage{multirow}     
\usepackage{array}        
\usepackage{booktabs}
\usepackage{multicol}
\usepackage{microtype}
\microtypesetup{nopatch=footnote} 
\newrobustcmd*\blacka[1]{\tikz[baseline=(char.base)]{
            \node[shape=circle,draw,inner sep=1pt,fill,text=white,minimum size=1em] (char) {\textsf{\small a}};}}

\newrobustcmd*\blackb[1]{\tikz[baseline=(char.base)]{
            \node[shape=circle,draw,inner sep=1pt,fill,text=white,minimum size=1em] (char) {\textsf{\small b}};}}

\newrobustcmd*\blackc[1]{\tikz[baseline=(char.base)]{
            \node[shape=circle,draw,inner sep=1pt,fill,text=white,minimum size=1em] (char) {\textsf{\small c}};}}

\newrobustcmd*\blackd[1]{\tikz[baseline=(char.base)]{
            \node[shape=circle,draw,inner sep=1pt,fill,text=white,minimum size=1em] (char) {\textsf{\small d}};}}

\newrobustcmd*\blacke[1]{\tikz[baseline=(char.base)]{
            \node[shape=circle,draw,inner sep=1pt,fill,text=white,minimum size=1em] (char) {\textsf{\small e}};}}

\newrobustcmd*\blackf[1]{\tikz[baseline=(char.base)]{
            \node[shape=circle,draw,inner sep=1pt,fill,text=white,minimum size=1em] (char) {\textsf{\small f}};}}

\newrobustcmd*\blackg[1]{\tikz[baseline=(char.base)]{
            \node[shape=circle,draw,inner sep=1pt,fill,text=white,minimum size=1em] (char) {\textsf{\small g}};}}

\newcommand{\eg}{e.g.\xspace}
\newcommand{\ie}{i.e.\xspace}

\newrobustcmd*\circled[1]
{\tikz[baseline=(char.base)]{
            \node[shape=circle,draw,inner sep=1pt,fill,text=white,minimum size=1.2em] (char) {\textsf{\small #1}};}}

\usepackage{algorithmic}

\begin{document}

\title{MERE: Hardware-Software Co-Design for Masking Cache Miss Latency in Embedded Processors}

\author{Dean You}
\affiliation{%
  \institution{National Center of Technology Innovation for EDA, School of Integrated Circuits}
  \institution{Southeast University}
  \city{Nanjing}
  \country{People's Republic of China}
}

\author{Jieyu Jiang}
\affiliation{%
  \institution{Sun Yat-Sen University University}
  \city{Guangzhou}
  \country{People's Republic of China}
}

\author{Xiaoxuan Wang}
\affiliation{%
  \institution{National Center of Technology Innovation for EDA, School of Integrated Circuits}
  \institution{Southeast University}
  \city{Nanjing}
  \country{People's Republic of China}
}

\author{Yushu Du}
\affiliation{%
  \institution{National Center of Technology Innovation for EDA, School of Integrated Circuits}
  \institution{Southeast University}
  \city{Nanjing}
  \country{People's Republic of China}
}

\author{Zhihang Tan}
\affiliation{%
  \institution{National Center of Technology Innovation for EDA, School of Integrated Circuits}
  \institution{Southeast University}
  \city{Nanjing}
  \country{People's Republic of China}
}

\author{Wenbo Xu}
\affiliation{%
  \institution{Huazhong University of Science and Technology}
  \city{Wuhan}
  \country{People's Republic of China}
}

\author{Hui Wang}
\affiliation{%
  \institution{National Center of Technology Innovation for EDA, School of Integrated Circuits}
  \institution{Southeast University}
  \city{Nanjing}
  \country{People's Republic of China}
}

\author{Jiapeng Guan}
\affiliation{%
  \institution{Dalian University of Technology}
  \city{Nanjing}
  \country{People's Republic of China}
}

\author{Zhenyuan Wang}
\affiliation{%
  \institution{National Center of Technology Innovation for EDA, School of Integrated Circuits}
  \institution{Southeast University}
  \city{Nanjing}
  \country{People's Republic of China}
}

\author{Ran Wei}
\affiliation{%
  \institution{Lancaster University}
  \country{People's Republic of China}
}

\author{Shuai Zhao}
\affiliation{%
  \institution{Sun Yat-Sen University University}
  \city{Guangzhou}
  \country{People's Republic of China}
}

\author{Zhe Jiang}
\affiliation{%
  \institution{National Center of Technology Innovation for EDA, School of Integrated Circuits}
  \institution{Southeast University}
  \city{Nanjing}
  \country{People's Republic of China}
}
\email{zhe.jiang@seu.edu.cn}

\begin{abstract}

Runahead execution is a technique to mask memory latency caused by irregular memory accesses. By pre-executing the application code during occurrences of long-latency operations and prefetching anticipated cache-missed data into the cache hierarchy, runahead effectively masks memory latency for subsequent cache misses and achieves high prefetching accuracy; however, this technique has been limited to superscalar out-of-order and superscalar in-order cores. For implementation in scalar in-order cores, the challenges of area-/energy-constraint and severe cache contention remain. 

Here, we build the first full-stack system featuring runahead, \texttt{\name},  from SoC and a dedicated ISA to the OS and programming model. 
Through this deployment, we show that enabling runahead in scalar in-order cores is possible, with minimal area and power overheads, while still achieving high performance. 
By re-constructing the sequential runahead employing a hardware/software co-design approach, the system can be implemented on a mature processor and SoC.
Building on this, an adaptive runahead mechanism is proposed to mitigate the severe cache contention in scalar in-order cores. Combining this, we provide a comprehensive solution for embedded processors managing irregular workloads.
Our evaluation demonstrates that the proposed \texttt{\name} attains 93.5\% of a 2-wide out-of-order core's performance while constraining area and power overheads below 5\%, 
with the adaptive runahead mechanism delivering an additional 20.1\% performance gain through mitigating the severe cache contention issues.

\end{abstract}

\maketitle

\label{sc:ov.4}
\label{sc:Overview}

\section{Introduction}
\label{SC:Intro}
Driven by increasing demands for real-time performance and user privacy in modern computing applications, irregular workloads such as sparse machine learning and graph processing are not only executed on Out-of-Order (OoO) cores in data centers and desktop systems but also increasingly run on Scalar In-Order (Scalar-InO) cores within embedded devices for efficient local data processing~\cite{hua2023edge,st_ai_atedge,jiang2021hiart,jiang2022high,fafoutis2018extending,gharib2018safety,shi2022srcn3d,sun2020pointmoseg,chen2021moving,cheng2021s3net,sun2021rsn}.
For instance, edge devices and Internet of Things (IoT) platforms leverage specialised Scalar-InO cores, e.g., ARM Cortex-M52 and Cortex-M55~\cite{arm_cortex_m52,arm_cortex_m55}, to perform sparse machine learning inference tasks. However, these workloads typically exhibit irregular memory access patterns (see Sec.~\ref{sc:Preliminaries}.1), resulting in frequent cache misses, which significantly prolong off-chip memory access times and degrade overall system performance~\cite{ainsworth2017software, ainsworth2018event, wang2025nvr, zhou2020graph, bailey1991parallel, jasak2009openfoam, izumo2004timidity++, naithani2021vector}.

Toward this, non-sequential memory access patterns brought by modern workloads and conventional hardware prefetchers (e.g., stream~\cite{srinath2007feedback} and global-history buffers~\cite{srinivasan2004continual}) have proved increasingly inadequate;
Hence, considerable research has been devoted to runahead techniques~\cite{mutlu2003runahead, mutlu2005reusing,hashemi2015filtered,hashemi2016continuous,naithani2020precise,naithani2021vector,naithani2023decoupled,naithani2024decoupled,roelandts2024scalar}.
Runahead techniques pre-execute application code during occurrences of long-latency operations (i.e., runahead mode), where the processor frees pipeline resources and checkpoints the architectural register state, facilitating recovery after prefetching operations. 
Once the initial long-latency operation completes, the processor exits runahead mode, restores the checkpointed state, and resumes normal execution starting from the initial long-latency instruction.
By prefetching the anticipated cache-missed data into the cache, runahead effectively masks memory latency for subsequent cache misses and achieves high prefetch accuracy even in irregular workloads (up to 95\%)~\cite{hashemi2016continuous}. 

While the runahead technique intuitively appears promising for masking the latency caused by irregular memory access in Scalar-InO processors, we found two fundamental incompatibilities when we tried to build it at the real RTL level:

\noindent(i) Previous research demonstrated that runahead mechanisms add only about 0.5\%$\sim$2\% area and 26.5\% power overheads (primarily due to unbeneficial runahead durations) to a complex superscalar OoO core (e.g., ARM Cortex-A76~\cite{arm_cortex_a76}), however, the modern Scalar-InO cores (e.g., ARM Cortex-M3~\cite{arm_cortex_m3}) possess an order-of-magnitude smaller area/power-approximately 1\% of the OoO cores. 
This means that, even modest overheads severely compromise the feasibility of directly integrating traditional runahead approaches into Scalar-InO cores.

\begin{figure}[t]
  \centering
  \begin{minipage}[ht]{0.45\textwidth}
    \centering
    \includegraphics[width=\linewidth]{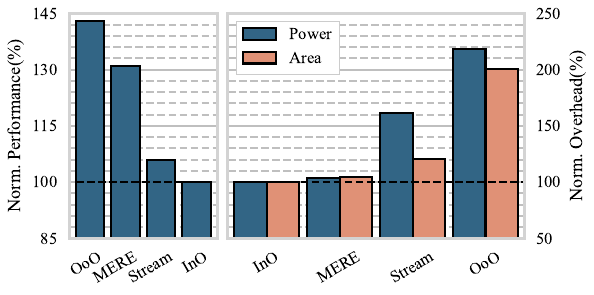}
    \caption{Average speedup (Norm.Performance) and whole-system power and area overheads for \name\ versus a scalar in-order baseline, stream prefetcher and an out-of-order core.}
    \Description{Average speedup (Norm.Performance) and whole-system power and area overheads for \name\ versus a scalar in-order baseline, stream prefetcher and an out-of-order core.}
    \label{fig:Comp}
  \end{minipage}
  \hfill
  \begin{minipage}[ht]{0.45\textwidth}
    \centering
    \includegraphics[width=\linewidth]{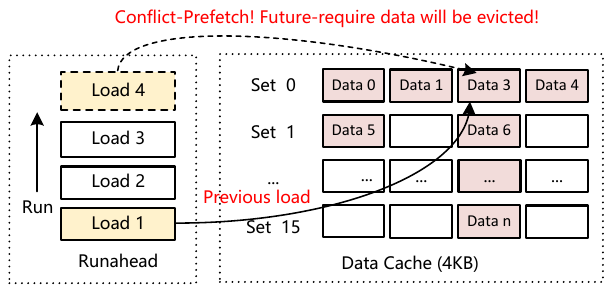}
    \caption{Confilct prefetch (The prefetch queue of runahead is located on the left, the cache state checkpoint is located on the right, load4 accesses data3, which will be used later) .}
    \Description{Comfilct prefetch (The prefetch queue of runahead is located on the left, the cache state checkpoint is located on the right, load4 accesses data3, which will be used later) .}
    \label{fig:comp-p}
  \end{minipage}
  \label{fig:intro1}
\end{figure}

\noindent(ii) We observed that runahead techniques, which heavily rely on  speculative execution, risk exacerbating cache pollution in Scalar-InO cores, due to the very limited cache capacity of the Scalar-InO cores.
When a conflict prefetch (when multiple prefetch addresses map to the same cache set, subsequent prefetches displace earlier prefetched blocks in the Data cache (D-Cache) ) occurs, a future-required data block from the D-cache may be evicted, thereby inducing cache contention (see Fig.~\ref{fig:comp-p}).
Unlike the OoO cores, Scalar-InO cores cannot tolerate extensive speculative memory operations without risking severe cache pollution and subsequent performance degradation.
To sum up, it is important but challenging to enable runahead in Scalar-InO cores, requiring a re-thinking of the methodology to manage irregular memory accesses effectively within resource-constrained Scalar-InO cores.

\begin{figure}[t]
    \centering
\includegraphics[width=1\columnwidth]{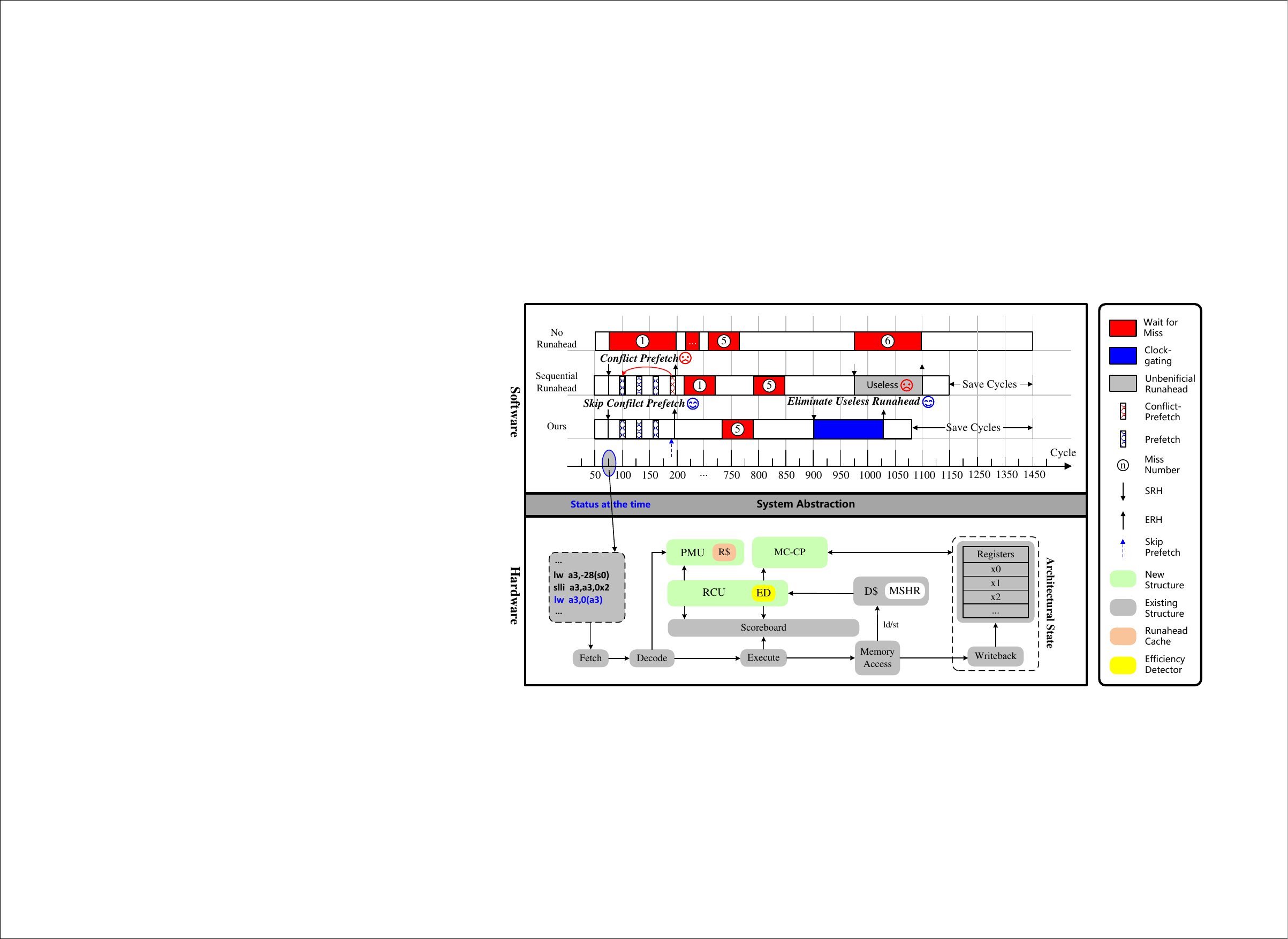}
    \caption{
    \name\ reconstructs the architecture of sequential runahead, software and hardware, In software, miss number 4 is conflict with miss number 1, and miss number 5 is an L1 miss, miss number 1-4/6 are L2 misses. 
    (SRH/ERH: Start/End Runahead; RCU: Runahead Control Unit; MC-CP: Multi-Cycle-CheckPoint; PMU: Prefetch Management Unit.)
    }
    \Description{\name\ reconstructs the architecture of sequential runahead, software and hardware, In software, miss number 4 is conflict with miss number 1, and miss number 5 is L1 miss, miss number 1-4/6 is L2 miss. 
    (SRH/ERH: Start/End Runahead; RCU: Runahead Control Unit; MC-CP: Multi-Cycle-CheckPoint; PMU: Prefetch Management Unit.)}
    \label{fig:Intro}
\end{figure}

\noindent \textbf{Contributions.}  
In this paper, we show that it is feasible to build runahead into a real Scalar-Ino core with minimal area/power overheads while achieving high performance. 
To do so, we reconstruct the sequential runahead, employing a hardware/software co-design approach, trading off functionalities across system layers.
This enables the runahead process to be precisely controlled, eliminates unbeneficial runahead duration, and allows conflict prefetch to be identified and skipped. 
We build a full-stack framework, \textbf{Make Each Runahead Effective (MERE)}, from the SoC and a dedicated Instruction Set Architecture (ISA) to the OS and programming model, providing a comprehensive solution for embedded processors managing irregular workloads.
We deployed the proposed system on an AMD Alveo U280 FPGA and evaluated it using a range of metrics, including overall throughput, prefetching performance, and overheads. The experimental results indicate that implementing the \name\ on Scalar-InO cores significantly improves the execution performance of irregular workloads. 
Our work achieves 93.5\% of the performance of a 2-wide OoO core while limiting area and power overheads to under 5\%, significantly outperforming superscalar OoO designs that incur double area and energy penalties (Fig.~\ref{fig:Comp}).
Moreover, with an average performance improvement of 20.1\% (Fig.~\ref{fig:makespan}), the adaptive runahead further enhances the performance of \name.

\section{Background and Related work}
\label{sc:Preliminaries}

In this section, we first discuss the background for the irregular memory access patterns (Sec.~\ref{sc:pre.1}), and existing runahead architectures on Scalar-InO, Super-Ino, and OoO cores (Sec.~\ref{sc:pre.2}). We summarise our proposed framework in Scalar-InO architecture with the prior works in Tab.~\ref{table:RW}.  

\subsection{Irregular memory access patterns}
\label{sc:pre.1}
\begin{figure*}[t]
    \centering
    \begin{minipage}[h]{0.54\textwidth}
        \begin{lstlisting}[language=C, label=lst:gcn, 
caption=Aggregation in Graph Convolution
, abovecaptionskip=0pt, 
belowcaptionskip=0pt
]
for (int x = 0; x < nnz * size; ++x) {
    i = x / size;
    j = x % size;
    temp = val[i]*feature[edge_col[i]][j];
    output[row[i]][j] += temp;
}
\end{lstlisting}
    \end{minipage}
    \hfill
    \begin{minipage}[h]{0.45\textwidth}
        \centering
        \includegraphics[width=\linewidth]{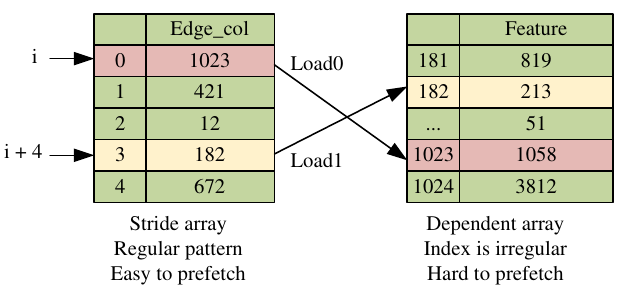}
        \caption{The process of indirct memory access.}
        \Description{The process of indirct memory access.}
        \label{fig:impfig}
    \end{minipage}
    \label{fig:imp}
\end{figure*}

Irregular memory access patterns are common for various workloads, particularly in fields like sparse machine learning~\cite{sliwa2020limits, xiao2019internet}, graph convolution networks~\cite{zhou2020graph,tan2022drips}, etc. 
Data associated with non-zero elements of sparse matrices or vectors is generally accessed indirectly. The process of accessing \texttt{feature[edge\_col[i]][j]}, which represents a feature vector, as illustrated in Listing 1, entails prefetching the column index array \texttt{edge\_col[i]}. This array corresponds to the column index of the i-th non-zero element in the feature matrix, enabling the relevant columns to be found. Index arrays \texttt{edge\_col[i]} frequently display the traits of irregular data structures and are typically static ( Fig.~\ref{fig:impfig} ), indicating that access to these arrays is usually sequential and can be easily captured by a stride prefetcher. 
Accessing the feature matrix through \texttt{feature[edge\_col[i]][j]} involves non-contiguous memory accesses. The large size of this matrix array, which cannot be fully cached, leads to numerous cache misses during indirect accesses.

For workloads that exhibit irregular memory access patterns, OoO cores can mask some of the memory latency by accommodating multiple loop iterations in the Reorder Buffer (ROB) simultaneously, with the extent of masking being dependent on the size of the ROB~\cite{arm_neoverse_v2,arm_cortex_a9}. By contrast, Scalar-InO cores have almost no tolerance, and even D-cache misses can significantly impact performance. Even when using a non-blocking cache (where the cores stall on use rather than on miss), the usage of miss-data will cause the core to halt execution until the long-latency main memory access is completed, leading to substantial performance loss~\cite{arm_cortex_m52,arm_cortex_m55,arm_cortex_m4}. Therefore, addressing memory latency issues is crucial for improving the performance of Scalar-InO cores.

\subsection{Existing Runahead Architectures}
\label{sc:pre.2}

Runahead techniques are applicable across multiple CPU architectures, including Scalar-InO, Super-InO, and OoO designs, as summarised in Tab.~\ref{table:RW}. 
Initially proposed by Dundas et al.~\cite{dundas1997improving} (basic runahead in Tab.\ref{table:RW}), this approach was first evaluated using a processor simulation that exclusively modelled the effects of data cache misses and subsequent prefetching behaviour.

In-order execution cannot tolerate any cache misses, as even with non-blocking caches, the pipeline inevitably stalls when the instructions require the pending miss-critical data, 
while out-of-order execution can mitigate some cache miss latency due to its ROB, allowing it to execute instructions that are independent of the cache miss. 
However, it is also unable to tolerate long-latency memory operations (e.g., last-level cache miss) due to the limited size of the ROB. Mutlu et al.~\cite{mutlu2003runahead} introduce Sequential Runahead (SR), which initially implements runahead in OoO cores. Instead of deploying a large, costly ROB, runahead alleviates performance decline induced by long-latency activities by pre-executing application code when long-latency operations occur. It also introduces the "runahead cache" to manage store/load instructions during runahead execution.   

\begin{table}[]
\caption{Prior runahead architectures.}

\label{table:RW}
{\tiny
\begin{tabular}{@{}c|c|cccccccc@{}}
\toprule
\multicolumn{1}{l|}{} &
  Property &
  \begin{tabular}[c]{@{}c@{}}Basic\cite{dundas1997improving}\\ Runahead\end{tabular} &
  \begin{tabular}[c]{@{}c@{}}SR\cite{mutlu2003runahead}\\ \end{tabular} &
  ERE\cite{mutlu2006efficient2} &
  PRE\cite{naithani2020precise} &
  VR\cite{naithani2021vector} &
  DVR\cite{naithani2023decoupled} &
  SVR\cite{roelandts2024scalar} &
  Ours \\ \midrule
\begin{tabular}[c]{@{}c@{}}Hardware\\ Architecture\end{tabular} &
  \begin{tabular}[c]{@{}c@{}}Core Type\\ Handle unbeneficial Runahead\\ Detect Indirct Memory Access\\ Resource-Constrain\\ Handle Cache-contention\\ Implementation method\end{tabular} &
  \begin{tabular}[c]{@{}c@{}}Scalar-InO\\ \ding{55}\\ \ding{55}\\ \checkmark\\ \ding{55}\\ Simulator\end{tabular} &
  \begin{tabular}[c]{@{}c@{}}OoO\\ \ding{55}\\ \ding{55}\\ \ding{55}\\ \ding{55}\\ Simulator\end{tabular} &
  \begin{tabular}[c]{@{}c@{}}OoO\\ \checkmark\\ \ding{55}\\ \ding{55}\\ \ding{55}\\ Simulator\end{tabular} &
  \begin{tabular}[c]{@{}c@{}}OoO\\ \checkmark\\ \ding{55}\\ \ding{55}\\ \ding{55}\\ Simulator\end{tabular} &
  \begin{tabular}[c]{@{}c@{}}OoO\\ \checkmark\\ \checkmark\\ \ding{55}\\ \ding{55}\\ Simulator\end{tabular} &
  \begin{tabular}[c]{@{}c@{}}OoO\\ \checkmark\\ \checkmark\\ \ding{55}\\ \ding{55}\\ Simulator\end{tabular} &
  \begin{tabular}[c]{@{}c@{}}Super-InO\\ \checkmark\\ \checkmark\\ \checkmark\\ \ding{55}\\ Simulator\end{tabular} &
  \begin{tabular}[c]{@{}c@{}}Scalar-InO\\ \checkmark\\ \checkmark\\ \checkmark\\ \checkmark\\ RTL\end{tabular} \\ \midrule
\begin{tabular}[c]{@{}c@{}}Programing\\ Model\end{tabular} &
  Software program method &
  \begin{tabular}[c]{@{}c@{}}hardware- \\
  only \end{tabular} &
  \begin{tabular}[c]{@{}c@{}}hardware- \\
  only \end{tabular} &
  \begin{tabular}[c]{@{}c@{}}hardware- \\
  only \end{tabular} &
  \begin{tabular}[c]{@{}c@{}}hardware- \\
  only \end{tabular} &
  \begin{tabular}[c]{@{}c@{}}hardware- \\
  only \end{tabular} &
  \begin{tabular}[c]{@{}c@{}}hardware- \\
  only \end{tabular} &
  \begin{tabular}[c]{@{}c@{}}hardware- \\
  only \end{tabular} &
  \begin{tabular}[c]{@{}c@{}}hardware/ \\software \\
  co-design \end{tabular} \\ \midrule
\begin{tabular}[c]{@{}c@{}}System\\ support\end{tabular} &
  \begin{tabular}[c]{@{}c@{}}Full SoC\\ OS support\end{tabular} &
  \begin{tabular}[c]{@{}c@{}}\ding{55}\\ \ding{55}\end{tabular} &
  \begin{tabular}[c]{@{}c@{}}\checkmark\\ \ding{55}\end{tabular} &
  \begin{tabular}[c]{@{}c@{}}\checkmark\\ \ding{55}\end{tabular} &
  \begin{tabular}[c]{@{}c@{}}\checkmark\\ \ding{55}\end{tabular} &
  \begin{tabular}[c]{@{}c@{}}\checkmark\\ \ding{55}\end{tabular} &
  \begin{tabular}[c]{@{}c@{}}\checkmark\\ \ding{55}\end{tabular} &
  \begin{tabular}[c]{@{}c@{}}\checkmark\\ \ding{55}\end{tabular} &
  \begin{tabular}[c]{@{}c@{}}\checkmark\\ \checkmark\end{tabular} \\ \bottomrule
\end{tabular}
}
\vspace{15pt}
\end{table}

\begin{figure}[t]
    \centering
    \begin{minipage}[t]{0.4\textwidth}
      \centering
      \includegraphics[width=\linewidth]{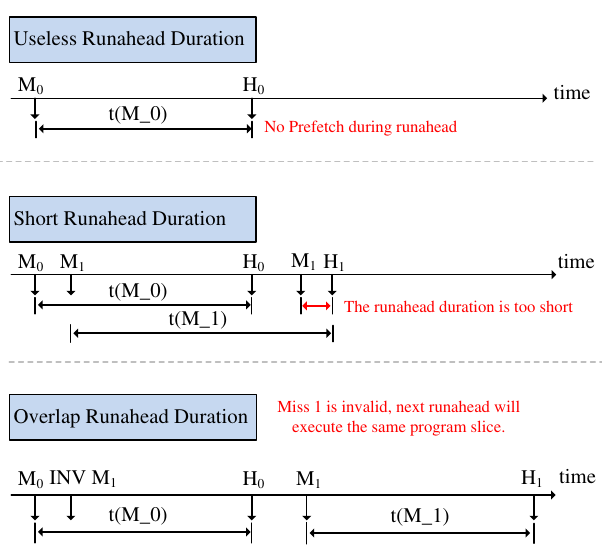}
      \caption{Three case of unbeneficial runahead. (INV $M_1$ means the load of miss number 1 is a invalid instruction, the defination of invalid instruction is in Sec.\ref{sc:RC2U}.)}
      \Description{Three case of unbeneficial runahead. (INV $M_1$ means the load of miss number 1 is a invalid instruction, the defination of invalid instruction is in Sec.\ref{sc:RC2U}.)}
      \label{fig:i_r}
    \end{minipage}
    \hfill
    \begin{minipage}[t]{0.55\textwidth}
      \centering
      \includegraphics[width=\linewidth]{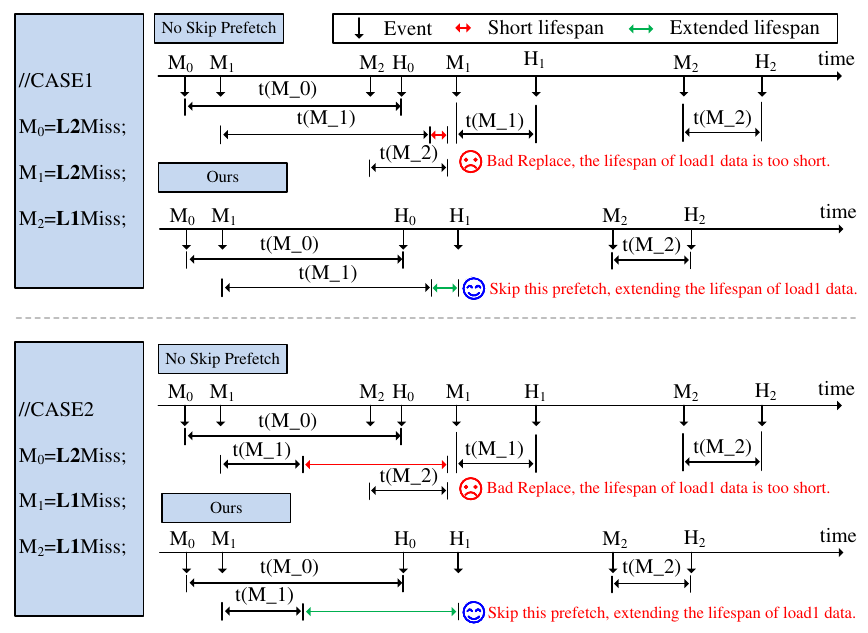}
      \caption{The case of conflict prefetch during runahead, where the addresses of \texttt{$M_1$} and \texttt{$M_2$}  are conflicted. ($M_n$ : miss number n; $H_n$ : hit number n; $t(M_n)$ : time of miss number n. )}
      \Description{The case of conflict prefetch during runahead, where the addresses of \texttt{load1} and \texttt{load2}  are conflicted. ($M_0$ : miss number n; $H_0$ : hit number n; $t(M_n)$ : time of miss number n. )}
      \label{fig:c_c}
    \end{minipage}
    \label{fig:contention}
\end{figure}

An unbeneficial runahead has three cases, and we show the specific description of them on Fig.\ref{fig:i_r}, including useless runahead (do not generate prefetch during runahead), short runahead (the runahead duration is too short), and overlap runahead (this runahead will execute the same program slice as the previous runahead, often caused by an invalid L2 miss).
This inefficiency stems from the non-negligible performance degradation and energy overheads incurred by pipeline flushing and refilling during enter/exit runahead mode. 
Such limitations persist in both basic runahead and sequential runahead implementations. 
To address these constraints, Mutlu et al.~\cite{mutlu2006efficient2} developed Efficient Runahead Execution (ERE) as an enhancement to sequential runahead methodologies.
The ERE introduces two key mechanisms:
(i) Runahead execution is triggered only when the memory access blocking operation has persisted for a predefined cycle threshold, ensuring the performance benefits outweigh transitional overheads.
(ii) Runahead duration is prohibited from overlapping with prior active runahead periods, eliminating redundant pipeline flushes.

Precise Runahead Execution (PRE)~\cite{naithani2020precise} augments standard runahead methodologies through three principal innovations:
(i) PRE exploits underutilised back-end microarchitectural resources (e.g., issue queue capacity and physical register file entries) to sustain speculative execution during runahead mode, thereby eliminating pipeline state flushing during mode transitions.
(ii) Instruction dispatch is constrained exclusively to load operations and their requisite address-generation dependencies following full instruction window stalls, minimising speculative overheads.
(iii) A hardware-guided mechanism rapidly recycles allocated back-end resources upon runahead termination, preserving structural integrity for non-speculative execution phases.
PRE's benefits originate from the processor's idle back-end resources and selective dispatch of load and address-generation instructions during runahead mode. 
However, PRE remains unable to prefetch most indirect memory accesses due to insufficient identification precision.

The SOTA runahead technique, Vector Runahead (VR)~\cite{naithani2021vector}, can generate high Memory Level Parallelsim (MLP) for indirect memory access patterns. It uses prediction tables to detect loads that indicate stride patterns. If these actions produce dependent loads within their computational sequence, several instruction chains will be created, and numerous subsequent iterations will be issued in parallel. 
Decoupled Vector Runahead (DVR)~\cite{naithani2023decoupled, naithani2024decoupled} is proposed as an enhancement to VR. Rather than triggering runahead upon the re-order buffer reaching capacity, it operates independently of the ROB size and autonomously issues speculative vectorised instruction streams, thereby enabling the processor to prefetch more extensively.
Both techniques show a significant capacity for masking memory latency. 
Unlike DVR, which uses spare physical registers for holding intermediate results of runahead execution, SVR utilises specified extra storage to retain intermediate outcomes of runahead execution (stores the scalar-vector instructions and the value of the speculative register file), ranging from 2KB to 9KB~\cite{roelandts2024scalar}. Even leveraging register reuse and reclamation strategies to minimise storage demands, this storage capacity remains considerably excessive in Super-InO cores and is even greater in Scalar-InO cores lacking superscalar pipelines.

\subsection{Research Challenge}
\label{sc:pre3}

Based on previous runahead research, we discover two primary challenges to implementation for integrating runahead in Scalar-InO cores: one is the hardware complexity and area/power constraints of mapping runahead from OoO to Scalar-InO cores; another is the prefetch conflict that is caused by cache contention in limited cache hierarchies of Scalar-InO cores. 

\textbf{(i) Hardware complexity and area/power constraints:}
In contrast to vector-series runahead techniques~\cite{naithani2021vector, naithani2023decoupled, roelandts2024scalar}, which demand N-way speculative register file/scalar-vector instruction replications for supporting vector-execution, sequential runahead~\cite{mutlu2003runahead} is more suitable for integration within Scalar-InO cores by requiring only single-context storage modules: the checkpoint and runahead cache.
However, the architectural gap between OoO and Scalar-InO cores fundamentally limits the direct transition of runahead microarchitecture, bringing significant design complexity that required a reconsideration of foundational modules and area/power optimisation strategies. 
For example, it is unclear how to perform pseudo-retirememt of instructions (the reasonable commit of instructions during runahead) without a ROB. 
This implies that we must intercept the writeback of all runahead invalid instructions (particularly load-miss instructions and their relative instructions) and collect extensive memory access messages and register operands (e.g., \texttt{load}/\texttt{store} addresses, memory access order and writeback tag).
Without ROB to maintain this execution metadata, 
it means that we need specific data extraction routes to gather these transient execution traces within limited timing windows before these data are overwritten. 
Besides the foundational modules, runahead integration in Scalar-InO cores requires careful consideration of introduced power overheads. 
While the modern fully hardware-managed approaches (ERE) for mitigating unbeneficial runahead durations (we show the three case of unbeneficial runahead duration in Fig.~\ref{fig:i_r}, which is proposed in~\cite{mutlu2006efficient2}) incur substantial hardware overheads and are incompatible with Scalar-InO architectures.
In summary, can we achieve the same benefit with minimal hardware and power overheads? 

\textbf{(ii) Prefetch conflicts.}
Furthermore, previous studies have overlooked the impact of cache contention in conjunction with different execution modes (see Fig.~\ref{fig:comp-p}), due to the large cache capacities in OoO architectures, which mitigate performance degradation.
However, the impact is particularly obvious in Scalar-InO cores with restricted cache hierarchies, where the cache capacity is very limited (e.g., in the ARM Cortex-M7~\cite{arm_cortex_m7} , which features a 4KB data cache with 4-way associativity, the number of cache sets is limited to only 16, making frequent evictions inevitable).
As our observation in Fig.~\ref{fig:c_c}, this critical issue arises from the tight coupling between rapid cache line replacements and short prefetched data lifespans, forcing high-priority data to be evicted prematurely before utilisation, which catastrophically impacts overall system performance. 
Fig.~\ref{fig:c_c} shows the performance bottleneck arising from the tight coupling between rapid cache line replacement and the short lifespan of prefetched data. 
This results in high-priority data being evicted prematurely -- before it can be used -- leading to severe performance degradation.
In Case1 of Fig.~\ref{fig:c_c}, $M_0$ is an L2 miss caused by an indirect memory access (marking the runahead duration between $M_0$ and $H_0$) , subsequent events $M_1$ (L2 miss) and $M_2$ (L1 miss) exhibit an address conflict, and $M_2$ has a higher prefetch priority. 
This conflict results in the data of $M_1$ being evicted by the $M_2$ (earlier response of L1 miss) before utilisation, 
leading to a short lifespan of the data of $M_1$ and generating an additional L1 miss penalty during normal execution.
The same pattern appears in Case2.
In summary, how can we reduce these penalties in Scalar-InO cores without introducing additional hardware complexity?
\section{\name: Overview}
\label{sc:ov}

In this section, we show how to build MERE in hardware and software, including the top-level concepts, the way to enable MERE in a mature processor and SoC, and ISA / programming model support.
As a demonstration purpose, we use a Scalar-InO core utilising the RISC-V ISA as an example, it features a five-stage pipeline, a non-blocking D-Cache and a 32-entry scoreboard. 
However, our work is not confined to this core and is general to modern Scalar-InO processor cores.

\subsection{Top-level Concepts}
\label{sc:ov.1}

\begin{figure*}[t]
    \centering
    \includegraphics[width=1\textwidth]{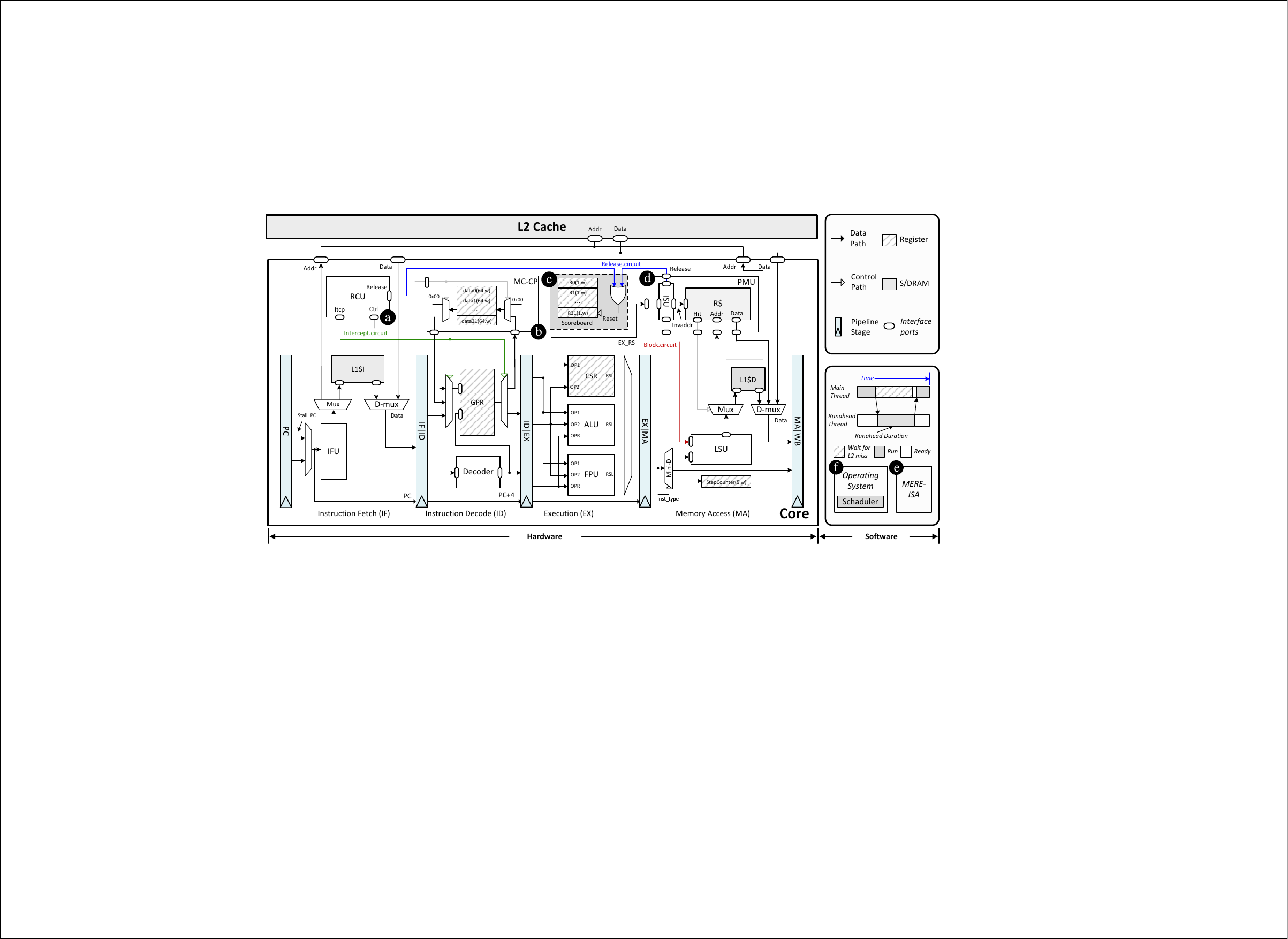}
    
    \caption{ An overview of \name. \textit{(Red: Block circuit; Blue:Release circuit; Green: Intercept circuit; CSR: Control Status Register; GPR: General-Purpose Register; ISU: Invalid Set Uint)}.
    At Hardware: \circled{a} RCU provides cycle-accurate control for the prefetching procedure;
    \circled{b} MC-CP saves and restores the processor state;
    \circled{c} Release.circuit releases the processor from the stall state;
    \circled{d} PMU detects and intercept erroneous prefetch requests; 
    At Software: \circled{e} Customised ISA, an abstraction layer for software-controlled interface. \circled{f} the OS scheduler, supporting adaptive runahead method.
    } 
    \Description{An overview of \name. \textit{(RCU: Runahead Control Unit; MC-CP: Multi-Cycle-CheckPoint; PMU: Prefetch Management Unit; CSR: Control Status Register; GPR: General-Purpose Register; ISU: Invalid Set Uint; R\$: Runahead Cache)}.
    At Hardware: \circled{a} RCU provides cycle-accurate control for the runahead procedure;
    \circled{b} MC-CP saves and restores the processor state;
    \circled{c} Release.Circuit releases the processor from the stall state;
    \circled{d} PMU detects and intercept erroneous prefetch requests; 
    At Software: \circled{e} Customised ISA, an abstraction layer for software-controlled interface. \circled{f} the OS scheduler, supporting adaptive runahead method.}
    \label{fig:Overview_2.0}
\end{figure*}

As the challenge we show above, to implement \name\ with minimal area and power overheads, while achieving high performance, it was necessary to make careful design-choice partitions between hardware and software. 

On one hand, we had no alternative but to create a dedicated data-extraction channel to ensure the proper functioning of runahead in Scalar InO cores. For load/store address, memory access sequence and writeback tag, we collect these messages from D-Cache MSHR, which holds the miss message of the processor (Sec.~\ref{sc:RC1U}). For register operands (e.g., source register number or destination register number), we collect these messages from the ex-stage of the processor. 
To ensure minimal area overheads, we must carefully consider the two storage modules of our design. 
One is for register state preservation, rather than directly employing single-cycle checkpoints typical in OoO core designs, we are adopting multi-cycle checkpoint mechanisms. 
By compromising marginal timing performance, we reach a significant optimisation in area and power. 
The other is the runahead cache, which handles load/store communication during runahead execution. 
We customise a compact cache (for our implementation, with 8 sets, 2 ways, and a 2-word block size) based on two observations regarding load/store operations during runahead: (i) infrequent store-load dependency chains, and (ii) spatial locality deficiency in data blocks.
Moreover, in contrast to earlier studies that rely on a fully hardware-managed strategy to mitigate unbeneficial runahead duration, our methodology simply incorporates a minimal StepCounter module (Fig.~\ref{fig:Overview_2.0}) (step defines when to terminate runahead based on prefetch benefits) with software-precomputed (Sec.~\ref{sc:theory3}) phase parameters. 

On the other hand, based on the constructed system, an adaptive runahead methodology (Sec.~\ref{sc:theory}) is proposed to (i) identify the conflict prefetch addresses that need to be skipped; (ii) determines the duration (i.e., the number of steps) of runahead. 
With this co-design philosophy, we achieve hardware frugality, displacing bulkier dedicated circuitry with lightweight coordination logic. 
Concurrently, to support our adaptive runahead method while avoiding the high resource overheads of full hardware support (unsuitable for resource-constrained Scalar-InO cores), we proposed MERE-ISA (Sec.~\ref{sc:ov.3}) to mediate hardware-software interaction. Encapsulated within the OS, it enables adaptive runahead with only a few lines of code (Sec.~\ref{sc:ov.4}).

Here, we show an overview of \name\ in Fig.~\ref{fig:Overview_2.0}. 
To ensure normal execution when exiting runahead, we save and recover core states at the Instruction Decode (ID) stage, while redirecting Stall\_PC (the PC value entering runahead) at the Instruction Fetch (IF) stage. 
During runahead, we release the pipeline when misses occur, and invalidate the relevant registers and addresses, while blocking the identified inaccurate address. Release and invalidation circuits are set at the Execution (EX) stage and blocking circuits are set at the Memory Access (MA) stage (it is also the implementation of \texttt{m.skip\_prefetch}). 
Moreover, we deploy a compact cache to store the result of stores that occur in runahead at the MA stage rather than directly storing it in D-cache. To prevent errors in the core execution due to the writing back of prefetched data to GPR, we constructed an intercept circuit at the Write Back (WB) stage. 
We also designed a mini-decoder (Mini-D) at the MA stage to execute \name\ ISAs, supporting the configuration of \name's characteristics. The microarchitecture design details for \name\ are in Sec. \ref{Micro}. 
With the above, we established a real SoC (Sec.~\ref{sc:ov.2}) and expanded the conventional RISC-V ISA to offer a dedicated interface for adaptive runahead(Sec.~\ref{sc:ov.3}). 

\subsection{Enabling \name\ in a mature processor and SoC}

\label{sc:ov.2}
Fig.~\ref{fig:Overview_2.0} illustrates the integration of \name\ into an SoC featuring a five-stage pipeline. 
The architectural design concept is as follows: we designed the Runahead Control Unit (RCU), to efficiently control the prefetch process of \name, ensuring that unbeneficial runaheads will be eliminated and prefetching will not interfere with the normal execution of the program (Fig.~\ref{fig:Overview_2.0} \circled{a}).
Checkpoint extraction and write-back circuits are established in the register file, featuring an MC-CP (Fig.~\ref{fig:Overview_2.0} \circled{b}), extracting and writing back the processor's state information. 
For miss requests (including invalid miss requests identified during runahead), the register number subsequently using the missing data is detected, and the corresponding position on the scoreboard is reset to release the pipeline (Fig.~\ref{fig:Overview_2.0} \circled{c}). We designed a Prefetch Management Unit (PMU) to invalidate erroneous prefetches and enabled memory access instructions throughout the runahead process (Fig.~\ref{fig:Overview_2.0} \circled{d}). An Invalid-Set Unit (ISU) was developed to track the register number and miss address responsible for pipeline release, thus preventing erroneous prefetches. Additionally, a compact cache, called Runahead-Cache (R\$) was constructed to gather the stored values of stores during  runahead, ensuring the execution of memory instructions. 

\begin{wraptable}{r}{9.0cm}
\caption{\name\ ISAs. (m: machine mode, non-privileged level)}
\label{table:ISA}

{\footnotesize
{%
\begin{tabular}{l|l}
\bottomrule
\hline
\rowcolor[HTML]{EFEFEF} 
\textbf{Instruction}          & \textbf{Description}                              \\ \hline
\texttt{{\textbf{m.check\_mode}}} rd          & Check if processor is in runahead mode   \\
\texttt{{\textbf{m.check\_skip}}} rd          & Check if this prefetch address need to skip    \\
\texttt{\textbf{m.skip\_prefetch}} rs1            & Skip rs1 address prefetch in runahead mode          \\
\texttt{\textbf{m.set\_step}} rs1 & Set the StepCounter as rs1               \\
\texttt{\textbf{m.clear\_step}} rs1 & Clear the StepCounter as rs1               \\

 \hline
 \toprule
\end{tabular}
}}
\end{wraptable}

\subsection{ISA Support}
\label{sc:ov.3}

In software layer: a customised MERE ISA is deployed as a control interface between software and hardware(Fig.~\ref{fig:Overview_2.0} \circled{e}). 
At the hardware level, a Mini-Decoder is employed to separate the conventional RISC-V ISA from MERE ISA. 

To support adaptive runahead (Sec.~\ref{sc:theory3}) and reduce microarchitectural complexity, we developed a customised ISA as an abstraction layer for software-controlled interfaces (Tab.~\ref{table:ISA}). The \texttt{check\_mode()} instruction verifies whether the core is in runahead mode. 
This works alongside \texttt{set\_step()} to regulate runahead duration. 
A pair of instructions, \texttt{check\_skip()} and \texttt{skip\_prefetch()}, work in tandem to skip prefetches that risk evicting unaccessed prior prefetch data. 
Finally, \texttt{clear\_step()} resets the step counter upon exiting runahead mode. Due to their simplicity and controllability, these instructions are designed as non-privileged (run in machine mode) operations, executable without requiring OS system calls.
Additionally, we develop an adaptive runahead function that is encapsulated using the \name\ ISA and are integrated into the operating system, where its internal scheduler handles task scheduling(Fig.~\ref{fig:Overview_2.0} \circled{f}).

\begin{figure*}[t]
\centering

\parbox[][8cm][c]{0.5\linewidth}{
\begin{minipage}[t]{0.5\textwidth}
\begin{algorithm}[H]
\caption{\footnotesize{Context switch(blue: added code).}}
\label{al:SW-L}
\footnotesize
\SetAlgoLined
$\vartriangleright$ \text{\texttt{Scheduler}}\\
\SetKwFunction{FMain}{\normalfont Context\_Switch}
    \SetKwProg{Fn}{Function}{:}{}
    \Fn{\FMain{\normalfont task \textit{*current}, core \textit{core\_index}}}{
        \textbf{Kernel}.Intr(DISABLE);
        task *\textit{next} = NULL;\\
        \emph{/* switching current task to the next task */}\\
    \textbf{Kernel}.Context.save(\emph{current});\\
        \emph{next} = \textbf{Kernel}.Find\_next();\\
        \If {\normalfont (\textcolor{blue}{\emph{next$\rightarrow$Adaptive\_Runahead}})}{
           \textbf{Kernel}.Context.init(\emph{next});\\
        }
        \Else {
        \textbf{Kernel}.Context.restore(\emph{next});\\}
       \emph{current} = \emph{next};\\
       \textbf{Kernel}.Intr(ENABLE);\\
\textbf{Kernel}.Context.jalr(\emph{current$\rightarrow$pc});\\
}
\textbf{End Function} \leavevmode\\
\end{algorithm}
\vfill
\hfill
\end{minipage}
}
\hspace{8pt}
\parbox[][8cm][c]{0.45\linewidth}{
\begin{minipage}[t]{0.45\textwidth}
\begin{algorithm}[H]
\caption{\footnotesize{Adaptive runahead function.}}
\label{al:SW-A}
\footnotesize
$\vartriangleright$ \text{\texttt{Adaptive runahead thread}}\\
\SetKwFunction{FMain}{\normalfont Adaptive\_Runahead}
    \SetKwProg{Fn}{Function}{:}{}
    \Fn{\FMain{}}{
        \emph{/* Check if processor is in runahead */}\\
        \If {\normalfont (\textbf{\name}.m.check\_mode())}{
            \emph{/* Set the value of StepCounter to decide when to exit runahead */}\\
            \textcolor{blue}{\textbf{\name}.m.set\_step()};\\
            \emph{/* Check if this prefetch address is confilct */}\\
    \If{\normalfont(\textbf{\name}.m.check\_skip())}
            {\emph{/* If conflict, then block this prefetch */}\\
            \textcolor{blue}{\textbf{\name}.m.skip\_prefetch();}}
            }
        \Else {
        \textbf{\name}.m.clear\_step();\\}
}
\vspace{4pt}
\textbf{End Function} \leavevmode\\
\end{algorithm}
\vfill
\hfill
\end{minipage}
}
\end{figure*}
\subsection{Adaptive runahead and Its Programming Model}
\label{sc:ov.4}

We encapsulate the adaptive runahead function based on the new ISA, leveraging context-switch functions. 
With only a few lines of code added to the kernel, it enables adaptive runahead automatically based on the input step value array and conflict prefetch address array (obtained via offline simulator training), while retaining standard thread scheduling and context-switching capabilities.
When the processor detects conditions suitable for entering runahead, it transitions to privileged mode and invokes context-switching to schedule tasks, switching to a new adaptive runahead thread. 
The adaptive runahead thread is initialised alongside the normal thread by extending the application thread's main function through constructor and destructor functions (Al.~\ref{al:SW-L}, line 13).

\noindent {\textbf{Programming model.}}
Firstly, check if the processor is in runahead mode. If active, the processor will invoke \texttt{m.set\_step} to configure the StepCounter value, directing the runahead thread(Al.~\ref{al:SW-A}: line 4 - 6). During runahead, continually invoke \texttt{m.check\_skip()} to detect whether the current prefetch address would overwrite a prior prefetched address (whose corresponding data has not yet been accessed). If such a conflict is detected, call \texttt{m.skip\_prefetch()} to skip the prefetch(Al.~\ref{al:SW-A}: line 8 - 10). Lastly, if the processor exits runahead mode, invoke \texttt{m.clear\_step()} to reset the StepCounter value(Al.~\ref{al:SW-A}: line 12).
\section{\name: The Microarchitecture}
\label{Micro}

As discussed, implementing runahead requires hardware support for new functionalities, which can significantly impact the existing core and the overall SoC design. We chose the open-source SoC, Rocket Chip~\cite{asanovic2016rocket,bachrach2012chisel}, as the foundation for the \name\ microarchitecture. It includes a low-power Rocket core, which supports the open-source RV64GC RISC-V ISA. 
It features 
a non-blocking D-cache and a frontend with branch prediction capabilities. The modular design of Rocket Chip exemplifies the characteristics of modern SoCs. 
By demonstrating this approach with Rocket Chip, we show that it can be applied to other SoCs, enabling the implementation of \name\
in most scalar embedded devices with an acceptable level of engineering efforts and overheads.
The top-level concepts and integration of \name\ into an SoC are discussed  in Sec.~\ref{sc:ov.1} and Sec.~\ref{sc:ov.2}; here we explore the microarchitecture design details in depth. 

\subsection{The Runahead Control Unit}
\label{sc:RC1U}
\begin{figure}[t]
    \centering
    
    \includegraphics[width=1\columnwidth]{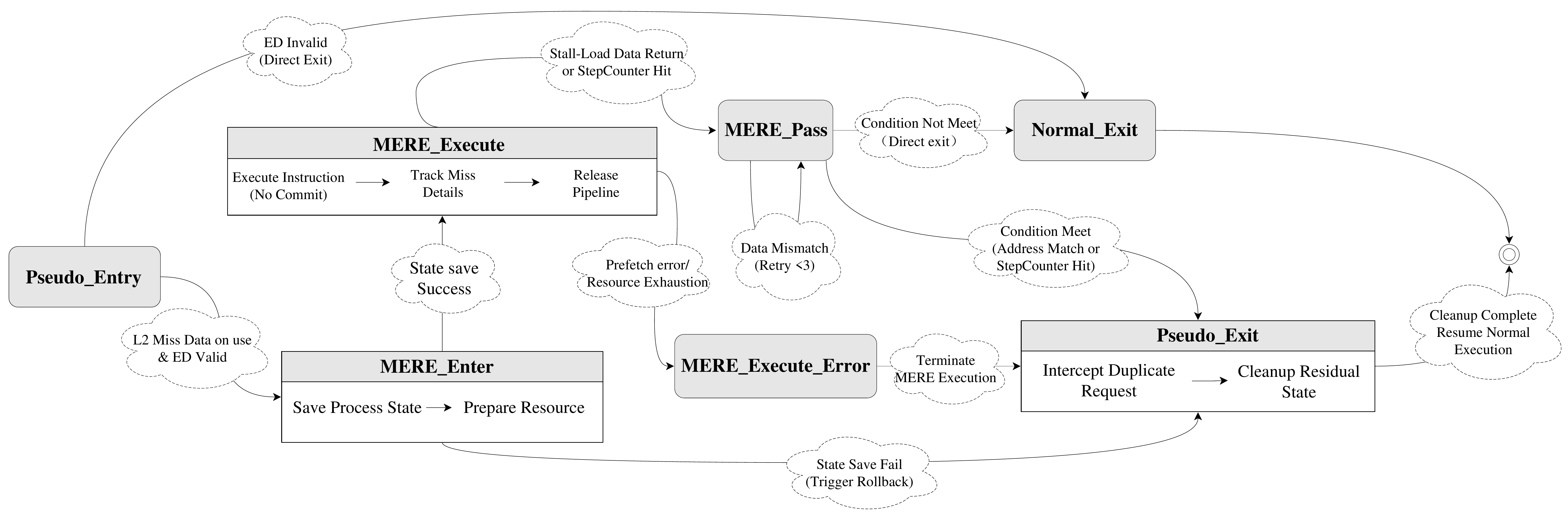}
    \caption{The Runahead FSM, a speculative recovery mechanism, upon detecting an L2 cache MSHR miss with sufficient resources, executes instructions without commit while tracking miss addresses, dynamically manages resources, and safely exits or rolls back based on data matching.
    \Description{The Runahead FSM, a speculative recovery mechanism, upon detecting an L2 cache MSHR miss with sufficient resources, executes instructions without commit while tracking miss addresses, dynamically manages resources, and safely exits or rolls back based on data matching.}
    }
    \label{fig:fsm}
\end{figure}

To eliminate unbeneficial runaheads and ensure normal execution on exiting runahead, an RCU is introduced. As the complex process of runahead and all the conditions during runahead can be counted, we integrated a Finite State Machine (FSM) with RCU(Fig.~\ref{fig:fsm}) to simplify design. 

\begin{figure}
\begin{minipage}[t]{0.48\textwidth}
    \centering
    
    \includegraphics[width=1\columnwidth]{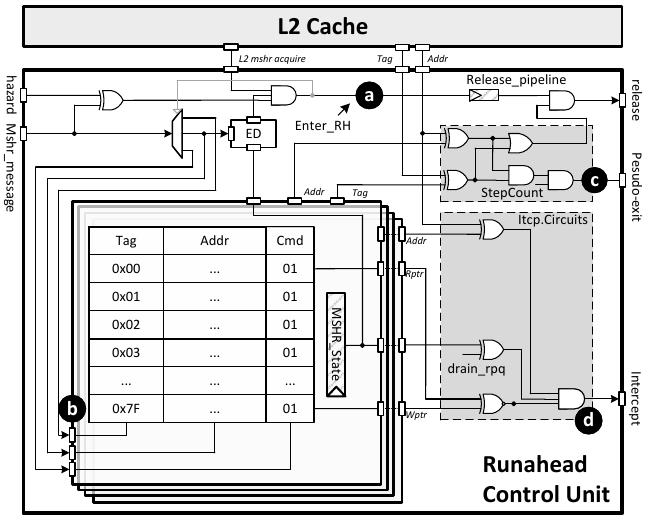}
    \caption{The Runahead Control Unit, controls the process of prefetch, 
    \circled{a} the condition of enter runahead; 
    \circled{b} MSHR trace the miss message of a stall-load or a gain-load; 
    \circled{c} the condition of entering pesudo-exit state is determined by StepCount or the stall-load is completed .
    }
    \Description{The Runahead Control Unit, controls the process of prefetch, 
    \circled{a} the condition of enter runahead; 
    \circled{b} MSHR trace the miss message of a stall-load or a gain-load; 
    \circled{c} the condition of entering pesudo-exit state is determined by StepCount or the stall-load is completed .}
    \label{fig:RC1U}
\end{minipage}
\hfill
\begin{minipage}[t]{0.48\textwidth}
    \centering
    \includegraphics[width=1\columnwidth]{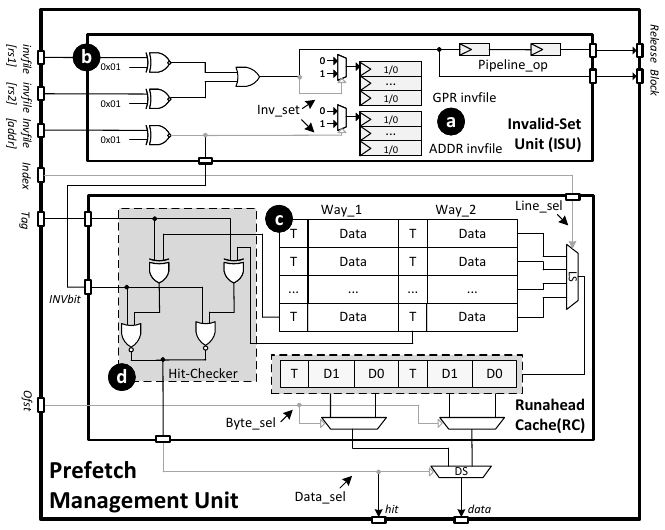}
    \caption{The Prefetch Management Unit  
    detects and blocks erroneous prefetch requests: 
    \circled{a} invfile records the invalid propagation of GPR and ADDR; 
    \circled{b} compares the RS and ADDR from EX stage with invfile; 
    \circled{c} a compact cache to store the results of stores in runahead. 
    }
    \Description{The Prefetch Management Unit  
    detects and blocks erroneous prefetch requests: 
    \circled{a} invfile records the invalid propagation of GPR and ADDR; 
    \circled{b} compares the RS and ADDR from EX stage with invfile; 
    \circled{c} a compact cache to store the results of stores in runahead.}
    \label{fig:RC2U}
\end{minipage}
\end{figure}

The FSM begins in a Pseudo\_Entry state, where it processes miss request information (write-back location) from the L2-cache MSHR. In this state,
the processor continues to execute, instead of stalling. 
The processor then uses the Efficiency Detector (ED) to identify whether this runahead is efficient. 
ED will acquire the load-miss address and the state bit of MSHR to ascertain whether this memory access is indirect and if the idle MSHR exceeds two. If these conditions are satisfied, 
the processor will transition to the MERE\_Enter state (Fig.~\ref{fig:RC1U} \circled{a}). 
This state initiates preparations for MERE\_Execute by saving the processor state, ensuring a smooth resumption of normal operations after MERE\_Execute.
On completing the tasks required in the MERE-Enter state, the processor will go directly to the MERE\_Execute state. 
In the MERE\_Execute state, the processor continues executing instructions without committing results to GPR, enabling effective prefetching and minimising idle time.
To facilitate this, the FSM tracks miss details (write-back register numbers, request addresses, and read/write pointers) of a stall-load or a gain-load  from the D-cache MSHR(Fig.~\ref{fig:RC1U} \circled{b}). 
Simultaneously, to prevent gain-loads from stalling the pipeline, the pipeline is released and the corresponding registers and addresses are invalidated by identifying the miss write-back register number and memory request address (for release detail see Sec.~\ref{micro:2}).If errors occur during this phase ,such as data mismatches (address conflicts), prefetch failures (invalid cache blocks), or resource exhaustion (idle MSHR $\leq$ 2),the FSM transitions to the MERE\_Execute\_Error state. 
Once the stall-load data returns or the StepCounter (the value of StepCounter is determined by \texttt{m.set\_step()}, see Sec.~\ref{sc:ov.3} and Sec.~\ref{sc:theory3}) hits, the FSM transitions to the MERE\_Pass state, which acts as an intermediary to determine whether the processor should move to a Pseudo\_Exit state or proceed directly to a Normal\_Exit state. In MERE\_Pass, address mismatches are re-evaluated through retries (Retry < 3). Successful retries loop back to MERE\_Execute; failures trigger a rollback or termination.
Two conditions allow the FSM to enter the Pseudo\_Exit phase: (i) if the benefit point is achieved before data write-back, by comparing the request address and read/write pointers, it is the basic terminate condition, or (ii) if the StepCounter reaches a specified value, signalling that the benefit point has been reached (Fig.~\ref{fig:RC1U} \circled{c}).
In the Pseudo\_Exit state, the FSM intercepts gain-loads related write-back requests by accurately detecting duplicate requests for both identical and different blocks. This interception prevents the \name{} process from being interrupted by replay mechanisms triggered by gain-loads in the same block. 
The FSM then finalises operations, ensuring all MERE\_Execute instructions are completed or safely discarded (Fig.~\ref{fig:RC1U} \circled{d}).

\subsection{The Multi-Cycle CheckPoint and Release Circuits}
\label{micro:2}

We constructed MC-CP to guarantee the core functions correctly when exiting runahead, with ``multi-cycle'' specifically for complex register file. We also created a specialised release circuit to flush the pipeline.

\noindent \textbf{Multi-Cycle CheckPoint:} The MC-CP, includes the Global History Register (GHR), Return Address Stack (RAS), and the GPR. The GHR and RAS handle branch history and return address tracking. When the core enters runahead, these structures are checkpointed ``in a single cycle'', preserving the information necessary for branch prediction and return address calculation. On exiting runahead, the saved branch history and return addresses are restored, maintaining accurate control flow without adding performance loss.
By contrast, the GPR, which stores the core architectural state, involves more data and complexity, leading to significant combinational logic costs and increased chip area overheads. To manage this, an MC-CP, which reduces the need for extensive module interfaces and lowers communication pressure across the core ought to be designed. 
Although checkpointing the architectural register file takes multiple cycles, it aligns with several cycles to clear and refill the pipeline when transitioning between runahead and normal modes, avoiding any additional performance penalties.

\noindent \textbf{Release Circuits:} Pipeline state management must interface directly with scoreboard-based control mechanisms in Scalar-InO cores. So, on identifying the usage of data from a stall-load or a gain-load, this structure receives the release signal from the RCU and PMU, resets the relevant register number in the scoreboard, and releases the processor from its stalled state.

\subsection{The Prefetch Management Unit}
\label{sc:RC2U}

We designed a PMU to detect and block erroneous prefetches while being able to handle memory access instructions during runahead. 

\noindent \textbf{Invalid-Set-Unit:} Similar to a scoreboard, the invfile records invalid registers and addresses. Each register or R\$ entry includes a bit indicating its validity (Fig.~\ref{fig:RC2U} \circled{a}). The destination register (RD) for a stall-load or a gain-load, as well as the invalid addresses stored during runahead, serve as sources for the invfile. 
We compare the source register (RS) from the EX stage and request addresses with the corresponding bits in the invfile, resulting in three scenarios (Fig.~\ref{fig:RC2U} \circled{b}): (i) When an RS number is present in the invfile, an invalid-propagation mechanism is initiated, setting the corresponding RD as invalid. (ii) If the load address is valid or all RSs are valid, an invalid-reset mechanism is triggered, resetting the invalid bit for the corresponding register. 
(iii) If a store address is found to be valid, an invalid-reset mechanism is activated, resetting the corresponding address bit. Based on the outcomes of these operations, the invalid signal is transformed into a blocking signal and a release signal for the core at the MA stage. 
This blocking signal will be transmitted to the Load Store Unit (LSU), preventing the request address from sending to memory hierarchy (applying \texttt{skip\_prefetch()} may support the interception of designated addresses, see Sec.~\ref{sc:ov.3}).
The invalid address is forwarded to the R\$ to ascertain whether the address is effectively hitting the R\$. 

\noindent \textbf{Runahead-Cache:} Due to the limited area resource of Scalar-InO cores, runahead cache requires parameter optimisation including block size and capacity tailoring.
In our implementation, the R\$ is designed as a compact two-way associative cache, with each entry containing a tag and data, where each data entry is two words (Fig.~\ref{fig:RC2U} \circled{c}). During runahead, the load accesses both R\$ and the D-cache simultaneously. It selects lines based on the index from the request address, matches the appropriate set, then selects bytes based on the offset, and finally retrieves the matched data according to the way hit. The hit mechanism involves comparing the tag of the request address with that of the R\$. If they match, it further verifies the data's validity. If valid, a hit signal is generated and used as the control signal for data selection (Fig.~\ref{fig:RC2U} \circled{d}). 

On exiting runahead, all values in the R\$ are invalidated to prevent access to outdated values until new runahead processes reset the stored addresses. Additionally, we adopted a pseudo-LRU policy to select the least recently used way for replacement.
\section{Enhancing Runahead in \name{} with Adaptive runahead}
\label{sc:theory}

With \name{} constructed, the runaheads are supported in Scalar-InO cores. 
However, existing runahead techniques (i) always prefetch each block of required data on cache regardless of whether useful data would be evicted and (ii) rely on a fixed termination condition (see Sec.~\ref{sc:RC1U}) without considering the cache state and memory accesses~\cite{mutlu2006efficient2, cain2010runahead, hashemi2015filtered, hashemi2016continuous, naithani2020precise}, leading to intensive cache contention with sub-optimal performance.
This section presents an optimisation method for runahead in \name{}, which decides (i) the duration and (ii) the beneficial prefetches for each runahead adaptively by exploiting the memory access sequence with runahead enabled. 
To achieve this, an analysis is constructed that estimates the memory access sequence with runahead enabled (Sec.~\ref{sc:theory2}). Then, the duration and the beneficial prefetches of each runahead are determined based on the analysis (Sec.~\ref{sc:theory3}). 

\subsection{System Model}
\label{sc:theory1}

We focus on a single Scalar-InO core equipped with a two-layered inclusive, non-blocking cache that has sufficient MSHR capacity.
Both L1 and L2 caches are set-associative with the Least Recently Used (LRU) applied as the replacement policy. 
The size of a cache line is 64 bit as with most of the modern in-order cores.
The number of ways is denoted $W_1$ and $W_2$, and the number of sets is denoted by $S_1$ and $S_2$, for both L1 and L2 caches respectively.
For memory accesses, their cache miss states are defined by $\Theta  = \{\texttt{L1\_HIT}, \texttt{L2\_HIT}, \texttt{L2\_MISS}\}$.
To obtain the cache status and the corresponding latency for a sequence of memory accesses, a two-layered cache simulator is constructed, which uses the LRU to update the cache given a set of memory accesses. 
The simulator can be configured with different settings (e.g., $W_1$ and $S_1$) and cache miss latency.
The implementation of a single-core LRU cache simulator (\eg, the classic cache in Gem5~\cite{binkert2011gem5}) is relatively straightforward and is omitted here.

The program has a sequence of memory accesses, denoted by $\Gamma=\{\tau_1,\tau_2,...\tau_n\}$. The execution time of $\tau_i$ is denoted by $C_i$, $C^\dagger_i$ and $C^\ddagger_i$ under the \texttt{L1\_HIT}, \texttt{L2\_HIT} and \texttt{L2\_MISS}, respectively. 
Function $\Theta(\tau_i) \in \Theta$ returns the cache miss state of $\tau_i$ based on the cache simulator. 
With the non-blocking cache applied, $\tau_i$ will be suspended and handled by the MSHR when it incurs an \texttt{L2MISS}, and the core continues to execute the following instructions that do not require $\tau_i$'s data.
The time from $\tau_i$'s execution to its data being demanded is denoted as $\delta_i$, which can be obtained from a timing analysis of the input program and the cache simulator.

If $\tau_i$'s data is not loaded when being required (\ie, after $\delta_i$ cycles from the execution of $\tau_i$), the system enters the runahead mode with a duration of $\lambda_i$, which finishes when $\tau_i$'s data arrives.
Function $F(\tau_i)$ denotes the sequence of memory accesses that are prefetched during this runahead, and $G(\tau_i)$ is the index of the memory access which triggers the runahead that prefetches $\tau_i$.
When the runahead finishes, the system switches back to normal mode to execute those instructions again with the preloaded data. The notations introduced by this section are summarised in Tab.~3.

\begin{figure}[t]
\centering
\includegraphics[width=0.7\columnwidth]{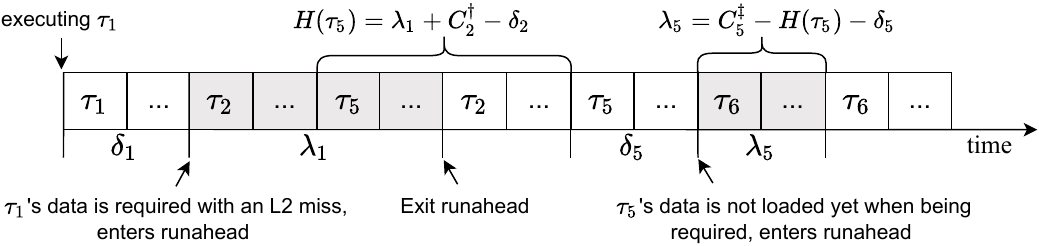}
\caption{An illustrative example of the execution with runaheads \textit{(blocks in white: normal execution; blocks in grey: runahead execution)}.}
\Description{An illustrative example of the execution with runaheads \textit{(blocks in white: normal execution; blocks in grey: runahead execution)}.}
\label{fig:Prefetch_time}
\end{figure}

Figure~\ref{fig:Prefetch_time} illustrates the system execution with runahead enabled.
In the example, $\tau_1$ triggers an \texttt{L2MISS} and is suspended. After $\delta_1$ cycles, the processor encounters an instruction dependent on the data from $\tau_1$ and enters runahead mode. The runahead phase lasts for $\lambda_5$ cycles, during which $\tau_5$ is executed to prefetch data. However, if $\tau_5$'s data is not yet loaded when required by the following instruction, the processor enters runahead mode again until the data becomes available.

\subsection{Analysing the Memory Access Sequence in Runaheads}
\label{sc:theory2}

To obtain the memory access sequence with runaheads, we compute the duration of each runahead (\ie, $\lambda_i$)  and the set of memory accesses being executed (\ie, $F(\tau_i)$). 
The $\lambda_i$ of each $\tau_i$ is computed in Eq.~\ref{eq:lambda}.
First, $\lambda_i = 0$ if $\tau_i$ is executed without an L2 miss. 
If an L2 miss occurs, a runahead is triggered after $\delta_i$ cycles when $\tau_i$ is executed (\ie, when $\tau_i$ data is required), and finishes when the data is obtained. 
In addition, it can be the case that $\tau_i$ was prefetched by a previous runahead, leading to a data loading time shorter than $C^\ddagger_i$. 
Thus, let $H(\tau_i)$ denote the period from the time that $\tau_i$ is prefetched to the time that it is executed in the normal mode, $\lambda_i$ is computed as $\max \{C^\ddagger_i-H(\tau_i)-\delta_i, 0\}$. 
For instance, the duration of the runahead triggered by $\tau_5$ is $\lambda_5=max\{C^\ddagger_5-\delta_5-H(\tau_5),0\}$ in Fig.~\ref{fig:Prefetch_time}.
\begin{equation} \label{eq:lambda}
\small
\lambda_i=
\begin{cases}
\max \{C^\ddagger_i-H(\tau_i)-\delta_i, 0\}  & \text { if~} \Theta(\tau_i)=\texttt{ L2\_MISS} \\
0 & \text { otherwise }
\end{cases}
\end{equation}

With $\lambda_i$ obtained, the set of memory accesses being executed in a runahead is computed in Eq.~\ref{eq:f}.
First, if $\lambda_i = 0$, then $F(\tau_i) = \varnothing$ as the runahead is not triggered. 
Otherwise (\ie, $\lambda_i > 0$), the runahead starts  $\delta_i$ cycles after $\tau_i$ is executed.
For a given $\tau_j$ with $i<j$, let $T_{i,j}$ denote the period from the start of $\tau_i$ to the start of $\tau_j$, $\tau_j$ will be prefetched by the runahead if $\delta_i < T_{i,j} \leq \delta_i+\lambda_i$. That is, the execution of $\tau_j$ is included in the runahead by its relative start time. Note, $T_{i,j}$ can be obtained based on the input program and the cache simulator.
\begin{equation} \label{eq:f}
\small
F(\tau_i)=
\begin{cases}
\left\{\tau_j \mid \delta_i < T_{i,j} \leq \delta_i+\lambda_i \right\} & \text {if~~} \lambda_i \neq 0 \\
\varnothing & \text {otherwise}
\end{cases}
\end{equation}

\begin{figure*}[t]
\centering

\hspace{-13pt}
\begin{minipage}[t]{0.47\textwidth}
\footnotesize
\setstretch{0.95}
\begin{algorithm}[H]
\For{\normalfont{each} $\tau_i \in \Gamma$}
{
\texttt{load}$(\tau_i)$;~~~~$F(\tau_i) \gets$ Eq.~\ref{eq:f};~~~~\\
\If{$F(\tau_i) \neq \varnothing$}{

    \texttt{load}$(\tau_j), \forall \tau_j \in [\tau_{i+1}, F(\tau_i).\text{head})$;\\
    $\vartriangleright$  \tiny
    {\scriptsize\ttfamily Find beneficial prefetches} \\
    \footnotesize
    \For{\normalfont{each} $\tau_j \in F(\tau_i)$}{
        \eIf{\normalfont{\texttt{!evict}}$(\tau_j)$}{
            \texttt{load}$(\tau_j)$;
        }{
            $F(\tau_i) = F(\tau_i) \setminus \{ \tau_{j} \}$;\\
        }
        
    }
    $\vartriangleright$ \tiny
    {\scriptsize\ttfamily Compute runahead duration} \\
    \footnotesize
    \While{$ T_{j,j+1} < {\normalfont\texttt{latency}}(\tau_{j+1})$}{
        \eIf{{\normalfont\texttt{!evict}}($\tau_{j+1}$)}{
            \texttt{load}$(\tau_j)$;\\
            $\lambda_i += T_{j,j+1}$;\\
            $F(\tau_i) = F(\tau_i) \cup \{ \tau_{j+1} \}$;~~~
            $j++$;\\
        } 
        {
            break;\\
        }
    }
    
}
}
\caption{Working process of the proposed adaptive runahead.}
\label{alg1}
\end{algorithm}
\end{minipage}
\hspace{8pt}
\begin{minipage}[t]{0.47\textwidth}
\vspace{-12pt}
\footnotesize
\caption*{\textbf{Table 3.} Notations introduced for constructing the adaptive runahead.}
\vspace{-14pt}
\renewcommand{\arraystretch}{1.045}
\begin{tabular}
{p{0.2\textwidth} p{0.75\textwidth}}
\label{tab:w_notations}
\tabularnewline \hline
\textbf{Notation} & \textbf{Description} \\
\hline
$W_1,S_1$ / & \multirow{2}{0.8\textwidth}{The number of cache ways and cache sets of the L1 and L2 cache, respectively.}\\
$W_2,S_2$ & \\
\hline
$\Gamma$ & A sequence of memory accesses required by a given program.\\
$\tau_i$ & The $i$\textsuperscript{th} memory access in $\Gamma$.\\
$T_{i,j}$& Time interval between $\tau_i$ and $\tau_j$.\\
$\theta_i$& Cache status (i.e., \texttt{L1HIT}, \texttt{L2HIT}, or \texttt{L2MISS}) of $\tau_i$ with runahead. \\
$C^{*}_i, C^{**}_i, C_i$ & The memory latency for cache status \texttt{L1HIT}, \texttt{L2HIT}, \texttt{L2MISS}, respectively.\\
\hline
$\delta_i$& The time duration from the finish of $\tau_i$ to the first use of its data.\\
$\lambda_i$ & Duration of the runahead triggered due to $\tau_i$.\\

$F(\tau_i)$ & Latest memory access that can be prefetched by runahead of $\tau_i$.\\
$G(\tau_i)$ & The index of earliest memory access in which the runahead fetches $\tau_i$.\\ 
$H(\tau_i)$ & The duration between the prefetching and the actual execution of $\tau_i$. \\

\hline
\end{tabular}
\end{minipage}
\vspace{8pt}
\end{figure*}

Finally, the duration between $\tau_i$ being prefetched and executed (\ie, $H(\tau_i)$) can be computed by Eq.~\ref{eq:h}.
The $H(\tau_i)$ consists of
(i) the time spent on the normal mode for executing the instructions in between and the runaheads triggered by the previous accesses, and (ii) the L1 cache miss latency incurred during normal execution. 
The first part can be computed by $\sum_{G(\tau_i)\leq j<i} \lambda_j$, where $G(\tau_i)= \max\left\{j \mid \tau_i \in F(\tau_j)\right\}$ gives the index of the memory access where its runahead fetches $\tau_i$.
We note that the normal execution within this duration is already accounted for in $\lambda_{G(\tau_i)}$, as it is also executed in the runahead of $\tau_{G(\tau_i)}$ before $\tau_i$ is prefetched.
The second part is computed as $\sum_{G(\tau_i)<j<i \wedge \Theta(\tau_j)=\texttt{L2\_HIT}} \max \{C_j^\dagger - \delta_j,0\}$ with the non-blocking time considered. The L2 cache miss would trigger runaheads that are considered in the first part, hence, are not considered. 
In addition, if $G(\tau_i)=\varnothing$, then $H(\tau_i) = 0$ as it is not prefetched by any previous runahead. The computation of $H(\tau_5)$ is illustrated in Fig.~\ref{fig:Prefetch_time}, assuming $\tau_2$ incurs an L1 cache miss when being executed.
\begin{equation}\label{eq:h}
\small
H(\tau_i)=\sum_{G(\tau_i)\leq j<i} \lambda_j + \sum_{\substack{G(\tau_i)<j<i \wedge\\\Theta(\tau_j)=\texttt{L2\_HIT}}} \max \{C_j^\dagger - \delta_j,0\}
\end{equation}

The above analysis computes $\lambda_i$ and $F(\tau_i)$ for every $\tau_i \in \Gamma$. The computation starts from $\tau_1$ with $G(\tau_1) = \varnothing$ and $H(\tau_1)=0$ so that $\lambda_1$ and $F(\tau_1)$ can be obtained directly. Then, the following accesses can processed based on $F(\cdot)$ of the previous ones.

\subsection{Adaptive runahead}
\label{sc:theory3}

Based on $F(\tau_i), \forall \tau_i \in \Gamma$, the adaptive runahead mechanism is constructed in Alg.~\ref{alg1}, which determines (i) the prefetches that should be performed and (ii) the duration of each runahead in the system by tracking the current cache state of the system. The following functions are implemented in the cache simulator to update its state: (i) \texttt{load($\tau_i$)} updates the cache by loading $\tau_i$'s data, (ii) \texttt{evict($\tau_i$)} returns whether a \texttt{load($\tau_i$)} would evict any prefetched data that have not been used, and (iii) \texttt{latency($\tau_i$)} returns the latency for loading $\tau_i$. 

For each $\tau_i \in \Gamma$ (starting from $\tau_1$), the algorithm updates the cache state by \texttt{load$(\tau_i)$} and determines whether $\tau_i$ can trigger a runahead based on $F(\tau_i)$ (lines 2-3). If so, the cache is first updated by the accesses executed between $\tau_i$ and its runahead (line 4). Then, the algorithm examines every $\tau_j \in F(\tau_j)$ to identify the prefetches that would evict useful data (lines 6-10), at which point the instruction \texttt{m.skip\_prefetch}() (see the ISA in Tab.~\ref{table:ISA}) and the \texttt{Adaptive\_Runahead} function (see Sec.\ref{sc:ov.4}) are used to skip such prefetches, achieving adaptive runahead that 
reduces cache contention.

Instead of exiting the runahead, the algorithm then examines the following memory accesses to determine whether they can be prefetched, based on their cache latency (lines 12-18). If the time needed to  execute $\tau_{j+1}$ (\ie, $T_{j,j+1}$) is less than its latency , $\tau_{j+1}$ is prefetched with $\lambda_i$ and $F(\tau_i)$ updated accordingly (lines 13-16). 
Finally, the runahead terminates at line 18 with the $\lambda_i$ and $F(\tau_i)$ determined, where \texttt{m.set\_step}() is invoked to configure the runahead duration, realising the adaptive duration that further enhances the performance of \name{} by exploiting prefetching. 

As described, the proposed adaptive runahead requires the cache simulator and the analysis of the memory access sequence. The cache simulation and the analysis are performed offline to identify the memory accesses that should not be prefetched, determining the runahead duration. This would not impose significant overheads at runtime.
In practice, the cache configurations (e.g., line size, cache latency, and cache miss latency) of the underlying hardware are provided by users for the simulation. The time complexity of the cache simulation is $\mathcal{O}(n)$ and the working process of adaptive runahead (Alg. 1) is $\mathcal{O}(n^2)$, where only the memory access behaviours are simulated using a list without accessing actual data.

\begin{table}[t]
\centering
\begin{minipage}[t]{0.47\textwidth}
\centering
\tiny
\caption{Hardware configurations evaluated}
\label{tab:core config}
\begin{tabular}{lccc}
\hline
             & Scalar-InO            & Super-InO            & OoO             \\ \hline
Core &
  \begin{tabular}[c]{@{}c@{}}1-wide, @1GHz,\\ 5-stage\end{tabular} &
  \begin{tabular}[c]{@{}c@{}}2-wide, @1GHz,\\ 6-stage\end{tabular} &
  \begin{tabular}[c]{@{}c@{}}2-wide, @1GHz,\\ 10-stage\end{tabular} \\
\begin{tabular}[c]{@{}l@{}}Scoreboard\\ ROB\\ Load/Store queue\\ Issue queue\end{tabular} &
  \begin{tabular}[c]{@{}c@{}}32\\ —\\ —\\ —\end{tabular} &
  \begin{tabular}[c]{@{}c@{}}32\\ —\\ —\\ —\end{tabular} &
  \begin{tabular}[c]{@{}c@{}}—\\ 32\\ 12\\ 32\end{tabular} \\
Branch Pred. & G-Share               & G-Share              & TAGE            \\
L1 I-Cache   & \multicolumn{3}{c}{8KB, 4-way, 32-set}                         \\
L1 D-Cache   & \multicolumn{3}{c}{4KB, 4-way,16-set, 4MSHR,Stride prefetcher} \\
L2 Cache     & \multicolumn{3}{c}{64KB, 8-way, 8MSHR}                         \\
Memory \& OS & \multicolumn{3}{c}{4GB DDR3, @666MHZ \& Linux version 6.2.0}   \\ \hline
\end{tabular}
\end{minipage}
\hfill
\begin{minipage}[t]{0.45\textwidth}
\centering
\caption{workload configurations evaluated}
\tiny
\label{tab:workload config}
\begin{tabular}{p{.15\columnwidth}p{.4\columnwidth}p{.28\columnwidth}}
\hline
Benchmark & Source                 & Input                                                                \\ \hline
GCN      & GNN\cite{zhou2020graph}                     & \begin{tabular}[c]{@{}l@{}}Citeseer, \\ Cora, \\ Pubmed\end{tabular} \\
IntSort   & NAS\cite{bailey1991parallel}                    & Classes B                                                            \\
ConjGrad  & NAS\cite{bailey1991parallel}                    & Classes B                                                            \\
PMC       & OpenFOAM HPC~\cite{jasak2009openfoam}           & Cavity flow                                                          \\
LSV       & OpenFOAM HPC~\cite{jasak2009openfoam}           & Cavity flow                                                          \\
LSG       & OpenFOAM HPC~\cite{jasak2009openfoam}           & Cavity flow                                                           \\ 
Timidity  & Real World Application~\cite{izumo2004timidity++} & 1000000000                                                           \\ 
\hline
Simulator workload & Randomly synthesised memory access sequences & Number of accesses in $100k\sim 200k$ \\ \hline
\end{tabular}
\end{minipage}
\vspace{15pt}
\end{table}

\section{Experiment Evaluation}

\noindent \textbf{Experimental platform.} 
We implemented \name, OoO, Scalar-InO, Super-InO, and Stream  on the AMD Alveo U280 FPGA, utilising the Rocket~\cite{keller2013risc} for the Scalar-InO core, the BOOM~\cite{zhao2020sonicboom} for the OoO core, and the Shuttle for the Super-InO core. Each core was equipped with an independent 4-way 4KB D\$ with 4 MSHR and a 4-way 8KB I\$, along with a shared 64KB L2\$ and external memory (4GB@666MHz). The configurations of the Scalar-InO core, Supers-InO core, and OoO core are 5-stage single-issue (@1GHz), 6-stage dual-issue (@1GHz), and 10-stage dual-issue (@1GHz), respectively (for more configuration details, see Tab.~\ref{tab:core config}).

\noindent \textbf{Workload setup.} In real-world workload, We evaluated a diverse set of benchmarks (see Tab.~\ref{tab:workload config}) that display intricate memory access patterns and computational dependencies during execution. These benchmarks encompass graph convolution networks, databases, and high-performance computing (HPC) workloads. Specifically, we utilised graph convolution networks (GCN)~\cite{zhou2020graph}, involving the multiplication of sparse matrices and feature matrices used in graph algorithms, with inputs from Citeseer (CS), Cora (CR), and Pubmed (PB). Additionally, we incorporated conjugate gradient (NAS-CG) and integer sort (NAS-IS) benchmarks from the NAS parallel benchmark suite~\cite{bailey1991parallel}. From OpenFOAM's HPC workloads~\cite{jasak2009openfoam}, we included PrimitiveMeshCheck (PMC), LeastSquaresVectors (LSV), and LeastSquaresGrad (LSG). Finally, we also considered the real-world application Timidity~\cite{izumo2004timidity++}. In simulator workload, We use extensive synthesised workload for evaluating the adaptive runahead mechanism. 

\subsection{Performance Overheads}

\begin{figure*}[t]
    \centering
    \includegraphics[width=1\textwidth]{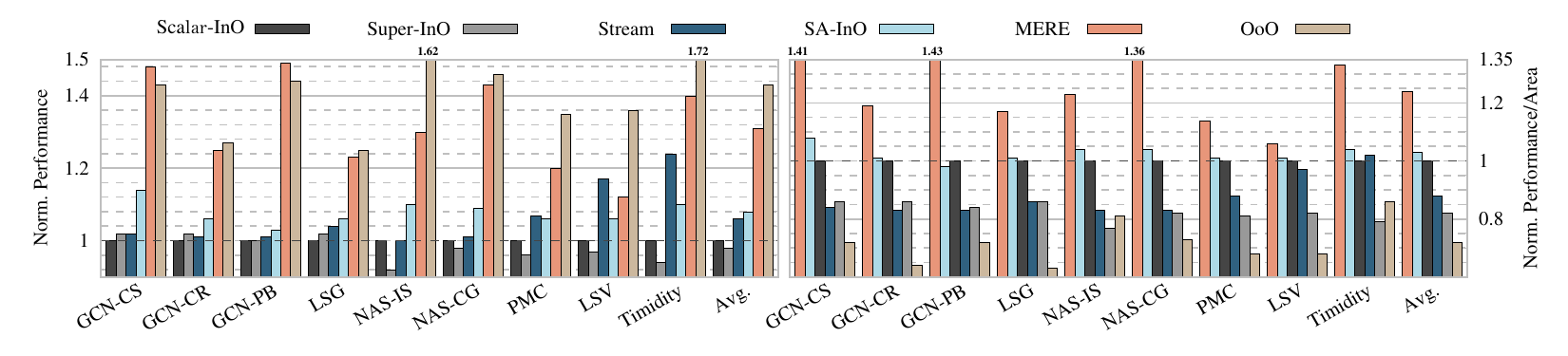}
    \vspace{-17pt}
    \caption{Normalised performance (left) and performance per area (right) for Scalar-InO, Super-InO, Stream, SA-InO, MERE, OoO (higher is better).}
    \Description{Normalised performance (left) and performance per area (right) for InO, Super-InO, Stream, SA-InO, MERE, OoO (higher is better).}
    \label{fig:Performance Overhead}
\end{figure*}

\noindent\textbf{Obs. 1.} In Fig.~\ref{fig:Performance Overhead}, \name\ demonstrated higher normalised performance, highlighting its architectural efficiency. This superior performance can be largely attributed to \name’s advanced ability to make better use of memory bandwidth.
Compared to the Same-Area InO processor (SA-InO), \name's advanced memory management and prefetching techniques offer more significant performance gains.
While SA-InO increases cache size to store more data close to the processor, this approach is less efficient than dynamically prefetching. As a result, SA-InO’s performance still falls short, particularly in workloads like graph computation tasks, which are memory-intensive and benefit significantly from \name’s ability to anticipate and prefetch data.

\noindent \textbf{Obs. 2.} Super-InO, Stream and SA-InO in Fig.~\ref{fig:Performance Overhead} show only marginal improvements over the baseline (Scalar-InO) in several workloads, and all significantly underperform \name\ in terms of normalised performance.
\name\ and OoO architectures, perform strongly in these workloads. Despite \name\ being based on a Scalar-InO core and an OoO core, their normalised performance is fairly comparable, with both significantly outperforming the baseline.
It is worth noting that Super-InO underperforms the baseline in some workloads, like NAS-IS, NAS-CG and PMC, as it uses the unoptimised Shuttle core, a RISC-V design not tailored for performance, area, or power efficiency.

In terms of performance per area, Super-InO and Stream perform poorly, often falling below the 
baseline. 
Despite slight gains in raw performance, these architectures fail to efficiently utilise chip area, highlighting their inefficiency in resource usage.
\name\ excels in both performance and area utilisation, making it ideal for area- and power-constrained systems where every inch of increased area requires corresponding performance gains to justify its value. Based on a Scalar-InO core, its efficient memory prefetching significantly reduces cache miss bottlenecks, achieving an average performance per area ratio of 1.24. 
By contrast, OoO, although offering higher raw performance through out-of-order execution, requires significantly more chip area, resulting in lower area efficiency. The increased silicon overheads of OoO’s complexity diminish some of its performance benefits. OoO’s normalised performance per area lags behind that of \name, making it less suitable for designs where chip area is a limited resource.

\subsection{Performance of adaptive runahead}
\label{sec:exptheory}

\noindent \textbf{Experimental setup.} 
The above justifies the effectiveness of the proposed \name{} using the entire SoC on the FPGA with real-world programs.
This section evaluates the proposed adaptive runahead (denoted as \texttt{Ours}) in terms of the number of cycles using extensive synthesised workloads, with different memory access patterns and cache configurations, covering a much wider number of memory accesses ($100k\sim 200k$ for each workload as shown in Tab.~\ref{tab:workload config}).
The following methods were applied for comparison: (i) \texttt{BS}: integrated runahead into the core with the basic terminate condition and without skip prefetching in Sec.~\ref{sc:RC1U} and
(ii) \texttt{BS|S}: a simple improvement on \texttt{BS} which stops runahead if the next prefetch evicts useful data.
The address of every access is randomly generated in the range $[0, D]$, where $D \in [24,112]$KB is the data size of the workload.
For an access $\tau_i$, $\delta_i$ is obtained by generating a random number of instructions in $[0, I]$ with
$I\in[3,8]$.  The execution time of each instruction is randomly decided within $1\sim180$ cycles following a weighted uniform distribution.
The cache was configured using $W_1=4$, $S_1=16$, $W_2=16$ and $S_2=128$; with $C_i$, $C^\dagger_i$ and $C^\ddagger_i$ set to 2, 25 and 180 cycles for all $\tau_i$, as commonly observed in COST architectures~\cite{karandikar2018firesim}.
To account for overheads entering and exiting runaheads, 5 cycles are added to each runahead duration.
For a system setting, 500 workloads were evaluated under the competing methods. 

 \begin{figure*}[t]
\centering
\subfigure[with $I=6$, $S_1=16$ and varied $D$.] 
{\label{fig:theory1}
\resizebox{.3\textwidth}{\dimexpr\height+1.5pt}{\includegraphics[width=.30\textwidth]{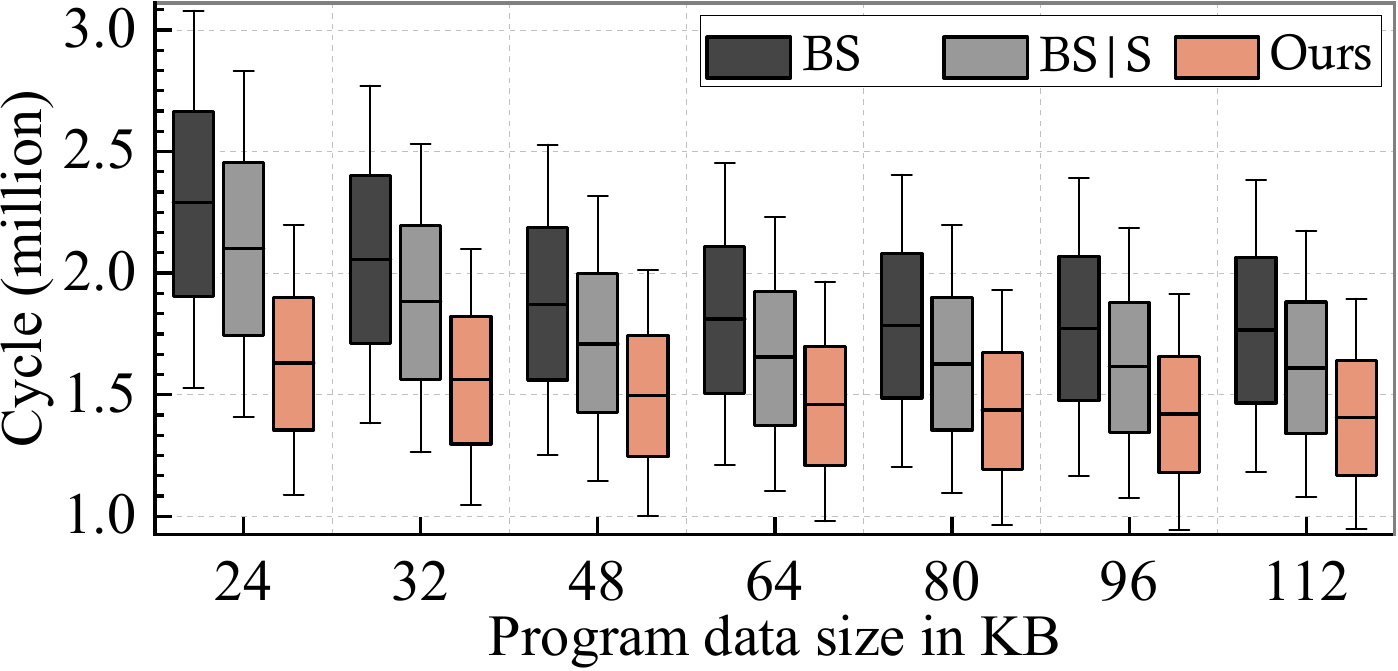}}
}
\centering
\hspace{0.0\textwidth}
\subfigure[with {$D= 32$, $S_1=16$ and varied $I$.}]{\label{fig:theory2}
\includegraphics[width=.30\textwidth]{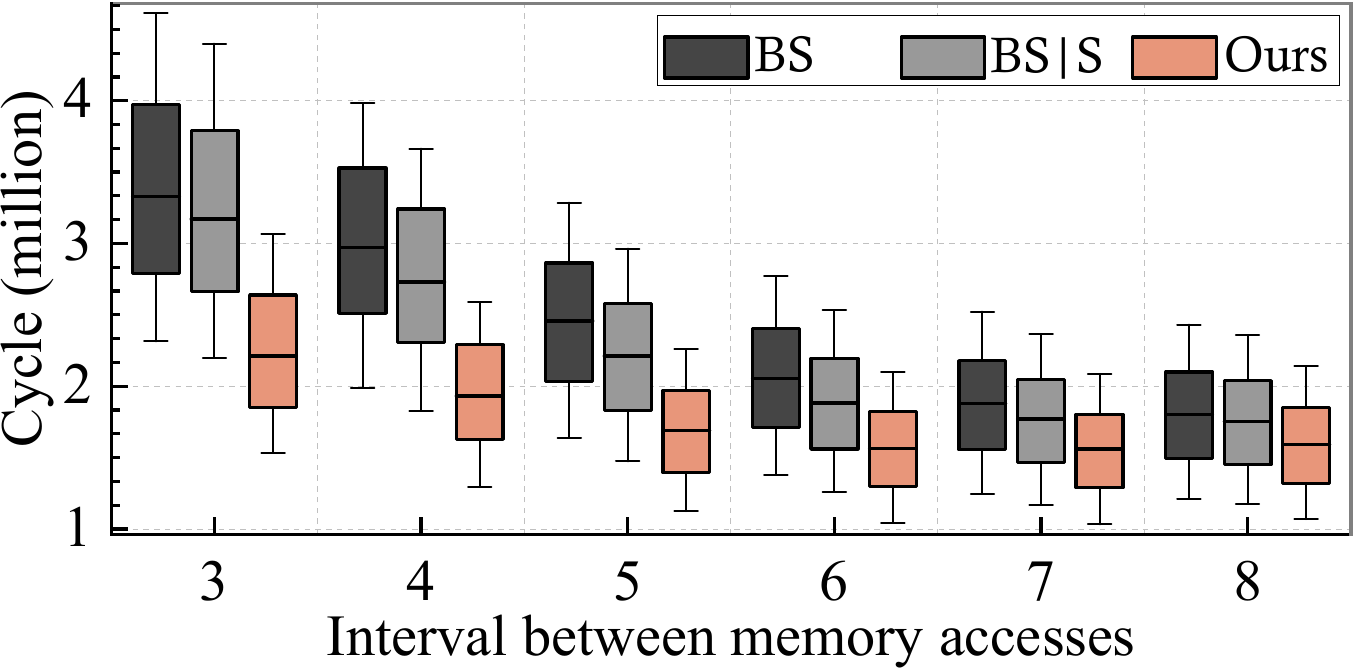}
}
\centering
\subfigure[with {$I=6$, $D = 32$ and varied $S_1$.}]{\label{fig:theory3}
\hspace{0.0\textwidth}
\includegraphics[width=.30\textwidth]{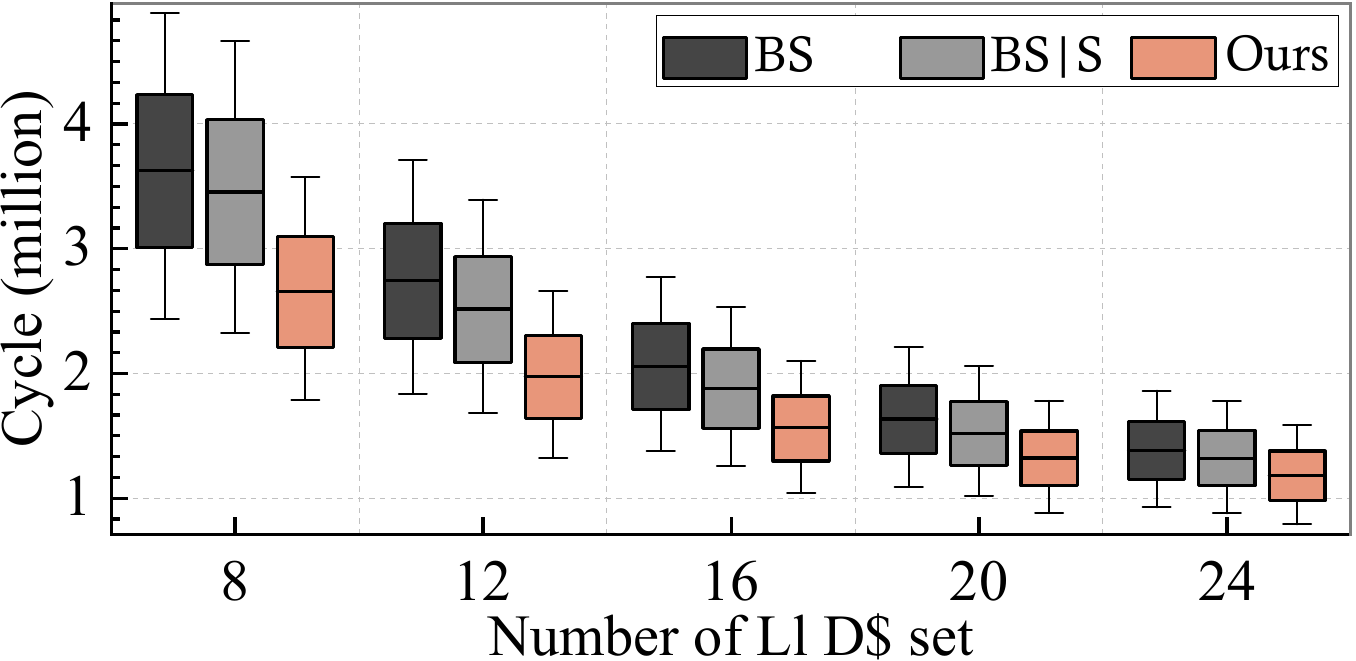}
}
\vspace{-7pt}
\caption{Performance comparison (in cycles) under varied $D$, $I$ and $S_1$.}
\Description{Performance comparison (in cycles) under varied $D$, $I$ and $S_1$.}
\label{fig:makespan}
\end{figure*}

\begin{figure}[t]
  \centering
  \begin{minipage}[t]{0.47\textwidth}
    \centering
    \includegraphics[width=\linewidth]{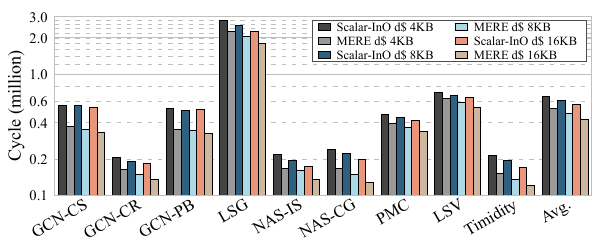}
    \caption{Impact of varying 4-way D-cache size on \name\ and baseline.}
    \Description{Impact of varying 4-way D-cache size on \name\ and baseline.}
    \label{fig:ways}
  \end{minipage}%
  \hfill
  \begin{minipage}[t]{0.47\textwidth}
    \centering
    \includegraphics[width=\linewidth]{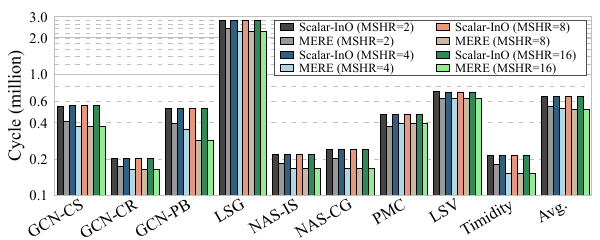}
    \caption{Impact of MSHRs on \name\ and baseline.}
    \Description{Impact of MSHRs on \name\ and baseline.}
    \label{fig:mshrs}
  \end{minipage}
  \label{fig:evlat1}
\end{figure}

\begin{figure}[t]
  \centering
  \begin{minipage}[t]{0.45\textwidth}
    \centering
    \includegraphics[width=\linewidth]{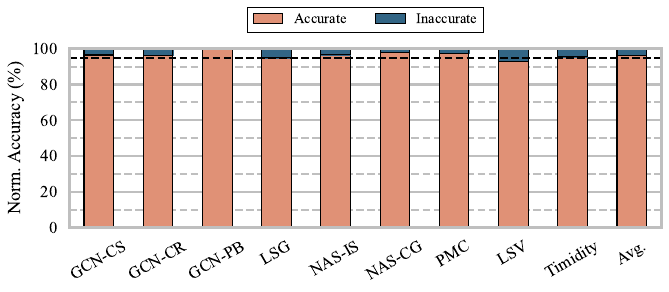}
    \caption{Prefetch accuracy of \name.}
    \Description{Prefetch accuracy of \name.}
    \label{table:hwo1}
  \end{minipage}%
  \hfill
  \begin{minipage}[t]{0.45\textwidth}
    \centering
    \includegraphics[width=\linewidth]{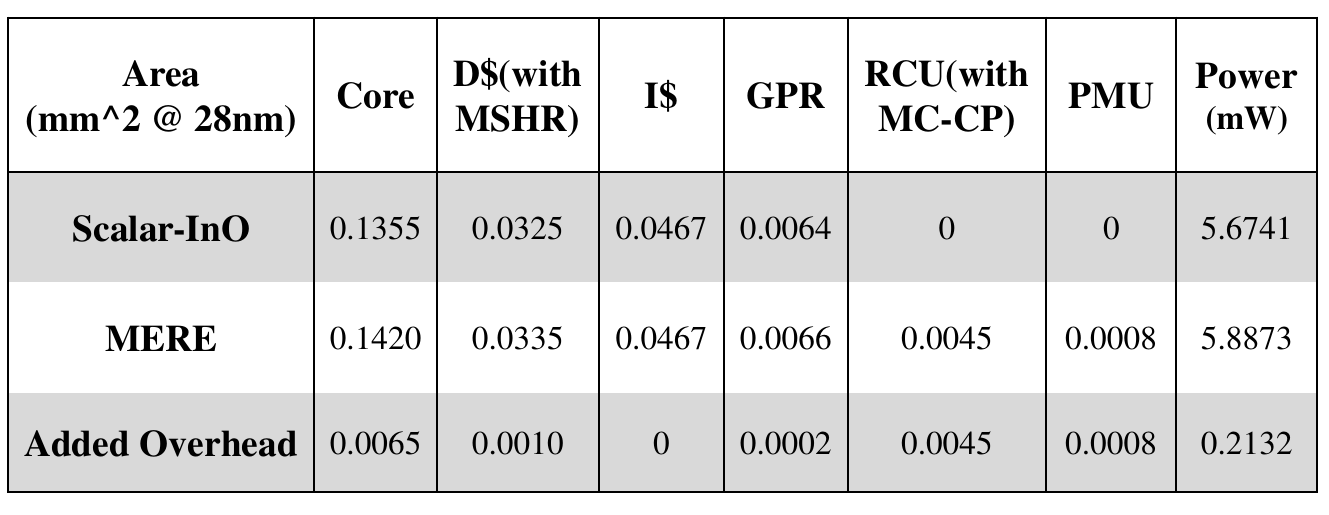}
    \caption{Hardware Overheads of \name.}
    \Description{Hardware Overheads (reported by DC).}
    \label{fig:accuracy}
  \end{minipage}
  \label{fig:evla2}
\end{figure}

\noindent \textbf{Obs. 3.}
The \texttt{Ours} outperformed both \texttt{BS} and \texttt{BS|S} in makespan. This can be observed from Fig.~\ref{fig:makespan}, in which it provided the lowest makespan in general, e.g., it outperformed \texttt{BS} by $20.1\%$ on average.  
In particular, the \texttt{Ours} showed a strong performance when $I \leq 6$ and $S_1 \leq 16$ in Fig.~\ref{fig:theory2} and Fig.~\ref{fig:theory3}, respectively. 
In such cases, the \texttt{BS} can cause frequent evictions of useful data, significantly increasing makespan due to intensive cache contention. Moreover, the \texttt{BS|S} showed observable improvements compared to \texttt{BS}, justifying the benefits of adaptive runahead by reducing cache contentions. This demonstrates that the traditional runahead can cause severe cache contention with undermined performance, and justifies the effectiveness of the proposed adaptive runahead method in a general case, especially when the cache is relatively small (\eg, when $S_1 \leq 16$ in Fig.~\ref{fig:theory3}).

\subsection{Sensitivity Analyses }

\noindent \textbf{Obs. 4.}
Fig.~\ref{fig:ways} illustrates the performance of \name\ and the Scalar-InO baseline under various D\$ sizes. The experimental findings indicate that the D\$ size affects the performance of baseline and \name\ to some extent when D\$ ways remain unchanged. For most workloads, the larger the D\$ size, the fewer cycles are required. With an average performance boost of 33\% above the baseline, the \name\ offers the most performance gain when the D\$ size is 16KB.  
Fig.~\ref{fig:mshrs} illustrates the variation in \name\ speedup with the increasing number of MSHRs in the D\$. \name\ is capable of accelerating the system only when the MSHRs are equal to or exceed 2, and it is constrained by the cache system's memory-level parallelism. Notably, \name\ achieves saturation at 8 MSHRs.

\subsection{Prefetch Accuracy}

\noindent \textbf{Obs. 5.}
We also observe the prefetch accuracy of the \name. \name\ sustains an accuracy beyond 95\% across most workloads, with an average accuracy of 96.4\%. Because the invalid prefetch requests, which are commonly due to index array fetch failures, will be intercepted by the PMU. 

\subsection{Hardware Overheads}

\noindent \textbf{Experimental setup.}
We synthesised a physical implementation of \name\ with Scalar-InO core baseline using TSMC 28nm PDKs~\cite{TSMC_28}.
The RTL was synthesised using Design Compiler (v2022.12), and the netlist was placed and routed via IC Compiler 2 (v2022.12), see Fig.~\ref{table:hwo1}.

\noindent \textbf{Obs. 6.}
The \name\ reported an area of 0.1420mm$^2$ and a power consumption of 5.8873mW, introducing only 0.0065mm$^2$ (4.8\%) of area and 0.2132mW (3.8\%) of power against the baseline. The increased area of the D\$ is part of the RCU combinational logic, while the increased area of the GPR is part of the checkpoint combinational logic. By adopting an MC-CP, this area has been significantly reduced.

\section{Conclusion}
This paper proposes the first full-stack system featuring runahead. 
This deployment demonstrates the possibility of transiting runahead in Scalar-InO cores.
By trading off architectural functionalities across hardware-software layers, MERE reconstructs sequential runahead microarchitecture to maintain area- and power-efficiency while achieving high performance.
Building up on this system, an adaptive runahead mechanism is introduced to mitigate the severe miss penalty that caused by cache contention in Scalar-InO cores. 
Experiments indicate that the proposed design has only a 10\% gap compared to a 2-wide OoO core in the performance of irregular workloads with both area and power overheads under 5\%. 
Moreover, our proposed adaptive runahead mechanism further enhances the performance by 20.1\%.  

\label{sc:Conclusion}

\bibliographystyle{ACM-Reference-Format}
\bibliography{sample-base}

\end{document}